\documentclass[12pt,preprint]{aastex}

\shorttitle{Spectroscopy of U Gem} \shortauthors{Echevarr\'\i{}a et
al.}

\begin{document}

\title{U Geminorum: a test case for orbital parameters determination}

\author{Juan Echevarr\'\i{}a\altaffilmark{1}, Eduardo de la Fuente\altaffilmark{2}
and Rafael Costero\altaffilmark{3}}
\affil{Instituto de Astronom\'\i{}a, {\it Universidad Nacional Aut\'onoma de M\'exico},\\
Apartado Postal 70-264, M\'exico, D.F., M\'exico}

\altaffiltext{1}{email: jer@astroscu.unam.mx}
\altaffiltext{2}{present address: Departamento de F\'\i{}sica,
CUCEI, Universidad de Guadalajara. Av. Revoluci\'on 1500 S/R
Guadalajara, Jalisco, Mexico.
                email: edfuente@astro.iam.udg.mx}
\altaffiltext{3}{email: costero@astroscu.unam.mx}

\begin{abstract}

High-resolution spectroscopy of \object{U Gem} was obtained during
quiescence. We did not find a hot spot or gas stream around the
outer boundaries of the accretion disk. Instead, we detected a
strong narrow emission near the location of the secondary star. We
measured the radial velocity curve from the wings of the
double-peaked H$\alpha$ emission line, and obtained  a
semi-amplitude value that is in excellent agreement with the
obtained from observations in the ultraviolet spectral region by
\citet{sio98}. We present also a new method to obtain $K_2$, which
enhances the detection of absorption or emission features arising in
the late-type companion. Our results are compared with published
values derived from the near-infrared NaI line doublet. From a
comparison of the TiO band with those of late type M stars, we find
that a best fit is obtained for a M6\,V star, contributing 5 percent
of the total light at that spectral region. Assuming that the radial
velocity semi-amplitudes reflect accurately the motion of the binary
components, then from our results: $ K_{em} = 107 \pm 2$  km
s$^{-1}$; $K_{abs} = 310 \pm 5$ km s$^{-1}$, and using the
inclination angle given by \citet{zha87}; $i = 69.7^\circ \pm 0.7$,
the system parameters become: $ M_{WD} = 1.20 \pm 0.05 \,
M_{\odot}$; $M_{RD} = 0.42 \pm 0.04 \, M_{\odot}$; and $ a = 1.55
\pm 0.02 \, R_{\odot}$. Based on the separation of the double
emission peaks, we calculate an outer disk radius of $R_{out}/a \sim
0.61 $, close to the distance of the inner Lagrangian point $L_1/a
\sim 0.63$. Therefore we suggest that, at the time of observations,
the accretion disk was filling the Roche-Lobe of the primary, and
that the matter leaving the $L_1$ point was colliding with the disc
directly, producing the hot spot at this location.

\end{abstract}


\section{Introduction} \label{intro}

Discovered by \citet{hin56}, U Geminorum is the prototype of a
subclass of dwarf novae, a descriptive term suggested by
\citet{pay38} due to the small scale similarity of the outbursts in
these objects to those of Novae.

After the work by \citet{kra62}, who found U~Gem to be a
single-lined spectroscopic binary with an orbital period around
4.25~hr, and from the studies by \citet{krz65}, who establish the
eclipsing nature of this binary, \citet{wan71} and \citet{sma71},
established the classical model for Cataclysmic Variable stars. The
model includes a white dwarf primary surrounded by a disc accreted
from a Roche-Lobe filling late-type secondary star. The stream of
material, coming through the L$_1$ point intersects the edge of the
disc producing a bright spot, which can contribute a large fraction
of the visual flux. The bright spot is observed as a strong hump in
the light curves of U Gem and precedes a partial eclipse of the
accretion disk and bright spot themselves (the white dwarf is not
eclipsed in this object).

A mean recurrence time for U Gem outbursts of $\approx$ 118 days,
with $\Delta$m$_V$=5 and outburst width of 12~d, was first found by
\citet{szk84}. However, recent analysis shows that the object has a
complex outburst behavior \citep{coo87, mat87, can02}.
\citet{sma04}, using the AAVSO data on the 1985 outburst, has
discovered the presence of super-humps, a fact that challenges the
current theories of super-outbursts and super-humps for long period
system with mass ratios above 1/3. The latter author also points out
the fact that calculations of the radius of the disc -- obtained
from the separation of the emission peaks \citep{kra75} in
quiescence -- are in disagreement with the calculations of the disc
radii obtained from the photometric eclipse data \citep{sma01}.

Several radial velocity studies have been conducted since the first
results published by \citet{kra62}. In the visible spectral range,
where the secondary star has not been detected, their results are
mainly based on spectroscopic radial velocity analysis of the
emission lines arising from the accretion disc \citep{kra62, sma76,
sto81, und06}. In other wavelengths, works are based on absorption
lines: in the near-infrared, on the Na I doublet from the secondary
star \citep{wad81, fri90, nay05} and in the ultraviolet, on lines
coming from the white dwarf itself \citep{sio98, lon99}.

Although the research work on U Gem has been of paramount importance
in our understanding of cataclysmic variables, the fact that it is a
partially-eclipsed and -- in the visual range -- a single-lined
spectroscopic binary, make the determination of its physical
parameters difficult to achieve through precise measurements of the
semi-amplitudes $K_{1,2}$ and of the inclination angle $i$ of the
orbit. Spectroscopic results of $K_{1,2}$  differ in the
ultraviolet, visual and infrared ranges. Therefore, auxiliary
assumptions have been used to derive its more fundamental parameters
\citep{sma01}. In this paper we present a value of $K_{1}$, obtained
from our high-dispersion Echelle spectra, which is in agreement with
the ultraviolet results, and of $K_{2}$ from a new method applicable
to optical spectroscopy. By chance, the system was observed at a
peculiar low state, when the classical hot spot was absent.

\section{Observations}

U Geminorum was observed in 1999, January 15 with the Echelle
spectrograph at the f/7.5 Cassegrain focus of the 2.1 m telescope of
the Observatorio Astr\'onomico Nacional at San Pedro M\'artir, B.C.,
M\'exico. A Thomson 2048$\times$2048 CCD was used to cover the
spectral range between $\lambda$5200 and $\lambda$9100 \AA, with
spectral resolution of R=18,000. An echellette grating of 150 l/mm,
with Blaze around 7000 \AA~, was used. The observations were
obtained at quiescence ($V \approx 14$), about 20 d after a broad
outburst (data provided by the AAVSO: www.aavso.org). The spectra
show a strong H$\alpha$ emission line. No absorption features were
detected from the secondary star. A first complete orbital cycle was
covered through twenty-one spectra, each with 10~min exposure time.
Thirteen further spectra were subsequently acquired with an exposure
of 5~min each. The latter cover an additional half orbital period.
The heliocentric mid-time of each observation is shown in column one
in Table~\ref{tab:RadVel}. The flux standard HR17520 and the late
spectral M star HR3950 were also observed on the same night. Data
reduction was carried out with the IRAF package\footnote {IRAF is
distributed by the National Optical Observatories, operated by the
Association of Universities for Research in Astronomy, Inc., under
cooperative agreement with the National Science Foundation.}. The
spectra were wavelength calibrated using a Th-Ar lamp and the
standard star was also used to properly subtract the telluric absorption
lines using the IRAF routine {\it telluric}.

\section{Radial Velocities}

In this section we derive radial velocities  from the prominent
H$\alpha$ emission line observed in U Gem, first by measuring the
peaks, secondly by using a method based on a cross-correlating
technique, and thirdly by using the standard double-Gaussian
technique designed to measure only the wings of the line. In the
case of the secondary star, we were unable to detect any single
absorption line in the individual spectra; therefore it was not
possible to use any standard method. However, here we propose and
use a new method, based on a co-adding technique, to derive the
semi-amplitude of the orbital radial velocity of the companion star.
In this section, we compare our results with published values for
both components in the binary. We first discuss the basic
mathematical method used here to derive the orbital parameters and
its limitation in the context of Cataclysmic Variables; then we
present our results for the orbital parameters -- calculated from
the different methods -- and finally discuss an improved ephemeris
for U Gem.

\subsection{Orbital Parameters Calculations} \label{orbparcal}

To find the orbital parameters of the components in a cataclysmic
variable -- in which no eccentricity is expected  \citep{zah66,
war95} -- we use an equation of the form
$$\mathrm{V(t)_{(em,abs)}} = \gamma + \mathrm{K_{(em,abs)}} sin[(2\pi(t - \mathrm{HJD_{\odot})/P_{orb}})],$$

where $\mathrm{V(t)_{(em,abs)}}$ are the observed radial velocities
as measured from the emission lines in the accretion disc or from
the absorption lines of the red star; $\gamma$ is the systemic
velocity; $\mathrm{K_{(em,abs)}}$ are the corresponding
semi-amplitudes derived from the radial velocity curve;
$\mathrm{HJD_{\odot}}$ is the heliocentric Julian time of the
inferior conjunction of the companion; and $\mathrm{P_{orb}}$ is the
orbital period of the binary.

A minimum least-squares sinusoidal fit is run, which uses initial
values for the four ($\mathrm{P_{orb}}$, $\gamma$,
$\mathrm{K_{em,abs}}$, and $\mathrm{HJD_{\odot}}$) orbital
parameters. The program allows for one or more of these variables to
be fixed, i.e. they can be set to constant values in the initial
parameters file.

If the orbital period is not previously known, a frequency search --
using a variety of methods for evenly- or unevenly-sampled time
series data \citep{sch99} -- may be applied to the measured radial
velocities in order to obtain an initial value for
$\mathrm{P_{orb}}$, which is then used in the minimum least-squares
sinusoidal fit. If the time coverage of the observations is not
sufficient or is uneven, period aliases may appear and their values
have to be  considered in the least-squares fits. A tentative
orbital period is selected by comparing the quality of each result.
In these cases, additional radial velocity observations should be
sought, until the true orbital period is found unequivocally. Time
series photometric observations are usually helpful to find orbital
modulations and are definitely important in establishing the orbital
period of eclipsing binaries. In the case of U Gem, the presence of
eclipses and the ample photometric coverage since the early work of
\citet{krz65}, has permitted to establish its orbital period with a
high degree of accuracy \citep{mar90}. Although in eclipsing
binaries a zero phase is also usually determined, in the case of
U~Gem the variable positions of the hot spot and stream, causes the
zero point to oscillate, as mentioned by the latter authors.
Accurate spectroscopic observations are necessary to correctly
establish the time when the secondary star is closest to Earth, i.e
in inferior conjunction. Further discussion on this subject is given
in section~\ref{ephem}.

To obtain the real semi-amplitudes of the binary, i.e
K$_{(em,abs)}$=K$_{(1,2)}$, some reasonable auxiliary assumptions
are made. First, that the measurements of the emission lines,
produced in the accretion disc, are free from distortions and
accurately follow the orbital motion of the unseen white dwarf.
Second, that the profiles of the measured absorption lines are
symmetric, which implies that the brightness at the surface of the
secondary star is the same for all its longitudes and latitudes.
Certainly, a hot spot in the disc or irradiation in the secondary
from the energy sources related to the primary will invalidate
either the first, the second, or both assumptions. Corrections may
be introduced if these effects are present. In the case of U Gem, a
three-body correction was introduced by \citet{sma76} in order to
account for the radial velocity distortion produced by the hot spot,
and a correction to heating effects on the face of the secondary
star facing the primary was applied by \citet{fri90} before equating
$\mathrm{K_{abs}} = \mathrm{K_{2}}$.

As initial values in our least-squared sinusoidal fits, we use
$\mathrm{P_{orb}} = 0.1769061911~d$ and $\mathrm{HJD_{\odot}} =
2,437,638.82325~d$ from \citet{mar90}, a systemic velocity of 42 km
s$^{-1}$ from \citet{sma01}, and K$_1 = 107$ km s$^{-1}$ and K$_2 =
295$ km s$^{-1}$ from \citet{lon99} and \citet{fri90}, respectively.
In our calculations, the orbital period was set fixed at the above
mentioned value, since our observations have a very limited time
coverage. This allow us to increase the precision for the other
three parameters.

\begin{table}
  \setlength{\tabcolsep}{0.8em} 
  \begin{center}
    \caption{Measured H$\alpha$ Radial Velocities.}
    \label{tab:RadVel}
    \begin{tabular}{llrrr}
       \hline
       \hline
       \noalign{\smallskip}
HJD            &      $\phi\tablenotemark{*}$ & Peaks\tablenotemark{a} & Fxc\tablenotemark{b} & Wings\tablenotemark{c} \\
(240000+)      &      & \multicolumn{3}{c}{(km s$^{-1}$)}   \\
       \tableline
       \noalign{\smallskip}
51193.67651 & 0.68 &     166.1   &  139.1   & 121.1 \\
51193.68697 & 0.75 &     183.4   &  130.0   & 133.8 \\
51193.69679 & 0.80 &     181.9   &  125.0   & 126.9 \\
51193.70723 & 0.86 &     167.9   &  102.0   & 101.1 \\
51193.71744 & 0.92 &     137.1   &   81.7   &  90.9 \\
51193.72726 & 0.97 &      90.0   &   46.8   &  41.7 \\
51193.73581 & 0.02 &      14.0   &  -17.9   &   6.9 \\
51193.74700 & 0.09 &     -47.9   &  -48.1   & -27.1 \\
51193.75691 & 0.14 &     -67.1   &  -66.7   & -48.2 \\
51193.76743 & 0.20 &     -99.6   &  -84.6   & -79.3 \\
51193.77738 & 0.26 &    -132.3   &  -86.1   &  -75.7 \\
51193.78900 & 0.32 &    -152.6   &  -60.2   &  -48.8 \\
51193.80174 & 0.39 &     -77.9   &  -32.9   &  -33.6 \\
51193.81211 & 0.45 &       9.0   &   10.9   &   14.5 \\
51193.82196 & 0.51 &     104.3   &   79.2   &   65.1 \\
51193.83176 & 0.56 &     134.6   &  113.7   &  107.0 \\
51193.84175 & 0.62 &     141.0   &  142.8   &  124.9 \\
51193.85156 & 0.67 &     159.3   &  158.6   &  147.6 \\
51193.86133 & 0.73 &     165.6   &  148.0   &  131.7 \\
51193.87101 & 0.79 &     192.9   &  142.8   &  130.3 \\
51193.88116 & 0.84 &     175.0   &  120.7   &  110.6 \\
51193.88306 & 0.91 &     154.6   &  106.5   &   91.1 \\
51193.90530 & 0.98 &      90.6   &   32.3   &   31.9 \\
51193.91751 & 0.05 &     -70.5   &    8.0   &  -23.1 \\
51193.93029 & 0.12 &     -88.5   &  -71.8   &  -51.6 \\
51193.94259 & 0.19 &     -97.1   &  -79.0   &  -66.7 \\
51193.95483 & 0.26 &    -114.4   &  -88.8   &  -75.6 \\
51193.95955 & 0.29 &    -142.2   &  -70.9   &  -67.9 \\

    \end{tabular}

\tablenotetext{*}{Orbital phases derived from the ephemeris given in
section~\ref{ephem}} \tablenotetext{a}{Velocities derived as
described in section~\ref{double-peaks}}
\tablenotetext{b}{Velocities derived as described in
section~\ref{template}} \tablenotetext{c}{Velocities derived as
described in section~\ref{hdgram}}

  \end{center}
  \end{table}

\begin{table}
  \setlength{\tabcolsep}{0.4em} 
  \begin{center}
    \caption{Orbital parameters derived from several radial
velocities calculations of the H$\alpha$ emission line.}
    \label{OrbParam}
    \begin{tabular}{lccc}
       \hline
       \hline
       \noalign{\smallskip}
Orbital & Peaks~(a) & Fxc~(b) & Wings~(c) \\
Parameters \\
       \tableline
       \noalign{\smallskip}
$\gamma$ (km s$^{-1}$)               &  38 $\pm$ 5    &  35 $\pm$ 3    & 34 $\pm$ 2     \\
$K$ (km s$^{-1}$)                    & 162 $\pm$ 7    & 119 $\pm$ 3    & 107 $\pm$ 2    \\
$\mathrm{HJD_{\odot}}$                & 0.8259(2)      & 0.82462(6)     & 0.82152(9)     \\
(+2437638 days) \\
$P_\mathrm{orb}$ (days)            &      (d)       &      (d)       &   (d)          \\
${\sigma}$                          &     25.2       &     12.2       &   9.1          \\
    \end{tabular}

\tablenotetext{a}{Derived from measurements of the double-peaks}
\tablenotetext{b}{Derived from cross correlation methods}
\tablenotetext{c}{Results from the fitting of fixed double gaussians
to the wings} \tablenotetext{d}{Period fixed, P=0.1769061911~d}

  \end{center}
  \end{table}

\subsection{The Primary Star} \label{prim}

In this section we compare three methods for determining the radial
velocity of the primary star, based on measurements of the H$\alpha$
emission line. Although, as we will see in the next subsections, the
last method results in far better accuracy and agrees with the
ultraviolet results, we have included all of them here because the
first method essentially provides an accurate way to determine the
separations of the blue and red peaks, which is an indicator of the
outer radius of the disc \citep{sma01}, and the second yields a
$K_{em}$ value much closer to that obtained from UV results than any
other published method. This cross-correlation method might be
worthwhile to consider for its use in other objects. Furthermore, as we will see in
the discussion, all three methods yield a consistent value of the systemic
velocity, which is essential to the understanding of other
parameters in the binary system.

To match the signal to noise ratio of the first twenty-one spectra,
we have co-added, in pairs, the thirteen 5-minute exposures. The
last three spectra were added to form two different spectra, in
order to avoid losing the last single spectrum. A handicap to this
approach is that, due to the large read-out time of the Thomson CCD,
we are effectively smearing the phase coverage of the co-added
spectra to nearly 900~s. However, the mean heliocentric time was
accordingly corrected for each sum. This adds to a total sample of
twenty-eight 600~s spectra.

\subsubsection{Measurements from the double-peaks} \label{double-peaks}

\begin{figure}[!]
  \begin{center}
     \includegraphics[width=0.9\columnwidth]{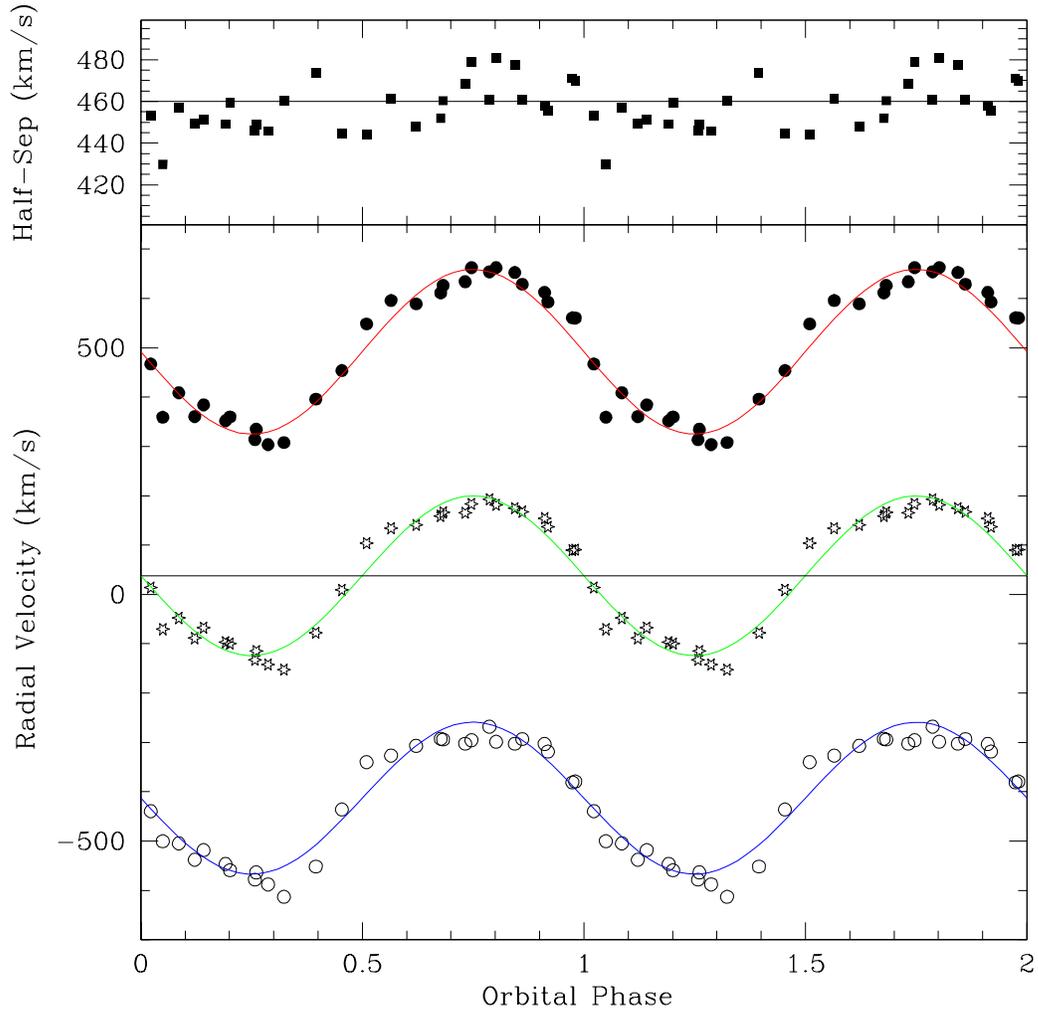}
     \caption{Radial velocity curve of the double peaks.
      The half-separation of the peaks, shown at the top of the diagram
      has a mean value of about 460 km s$^{-1}$. The curve at the middle
      is the mean from blue (bottom curve) and red (top curve).}
     \label{fig:dob-peak}
  \end{center}
\end{figure}

We have measured the position of the peaks using a double-gaussian
fit, with their separation, width and position as free parameters.
The results yield a mean half-peak separation $V_{out}$ of about
460~km~s$^{-1}$. The average value of the velocities of the red and
blue peaks, for each spectrum, is shown in column 3 of
Table~\ref{tab:RadVel}. We then applied our nonlinear least-squares
fit to these radial velocities. The obtained orbital parameters are
shown in column 2 of Table~\ref{OrbParam}. The numbers in
parentheses after the zero point results are the evaluated errors of
the last digit. We will use this notation for large numbers
throughout the paper. The radial velocities are also shown in
Figure~\ref{fig:dob-peak}, folded with the orbital period and the
time of inferior conjunction adopted in the section~\ref{ephem}. The
solid lines in this figure correspond to sinusoidal fits using the
derived parameters in our program. Although we have not
independently tabulated the measured velocities of the blue and red
peaks, they are shown in Figure~\ref{fig:dob-peak} together with
their average. The semi-amplitudes of the plotted curves are 154 km
s$^{-1}$ and 167 km s$^{-1}$ for the blue and red peaks,
respectively.

\subsubsection{Cross Correlation using a Template} \label{template}

\begin{figure}[!]
  \begin{center}
     \includegraphics[width=0.9\columnwidth,angle=90]{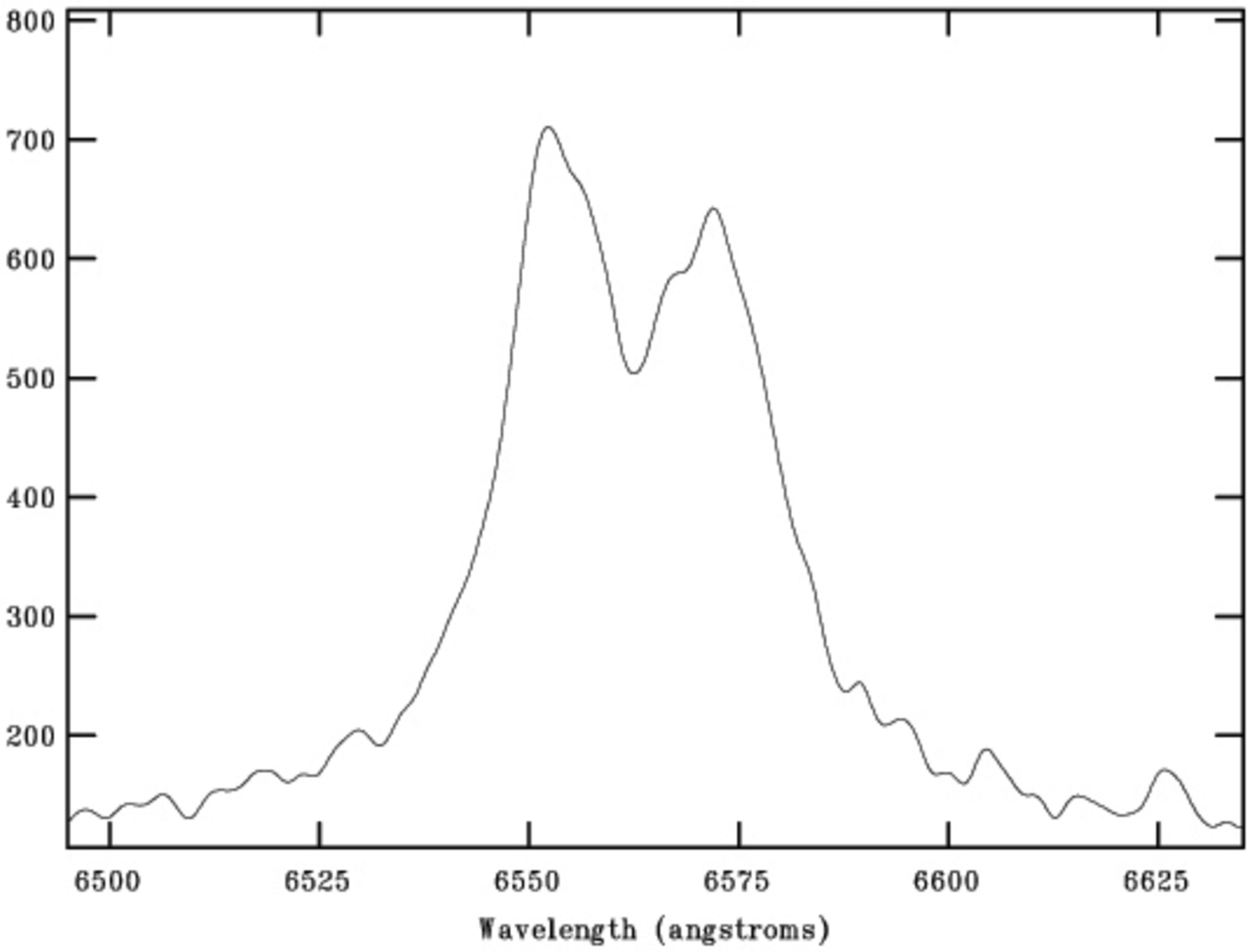}
     \caption{
H$\alpha$ template near phase 0.02. The half-separation of the peaks
has a value of 470 km~s$^{-1}$. }
     \label{fig:halfxc}
  \end{center}
\end{figure}

\begin{figure}[!]
  \begin{center}
     \includegraphics[width=\columnwidth]{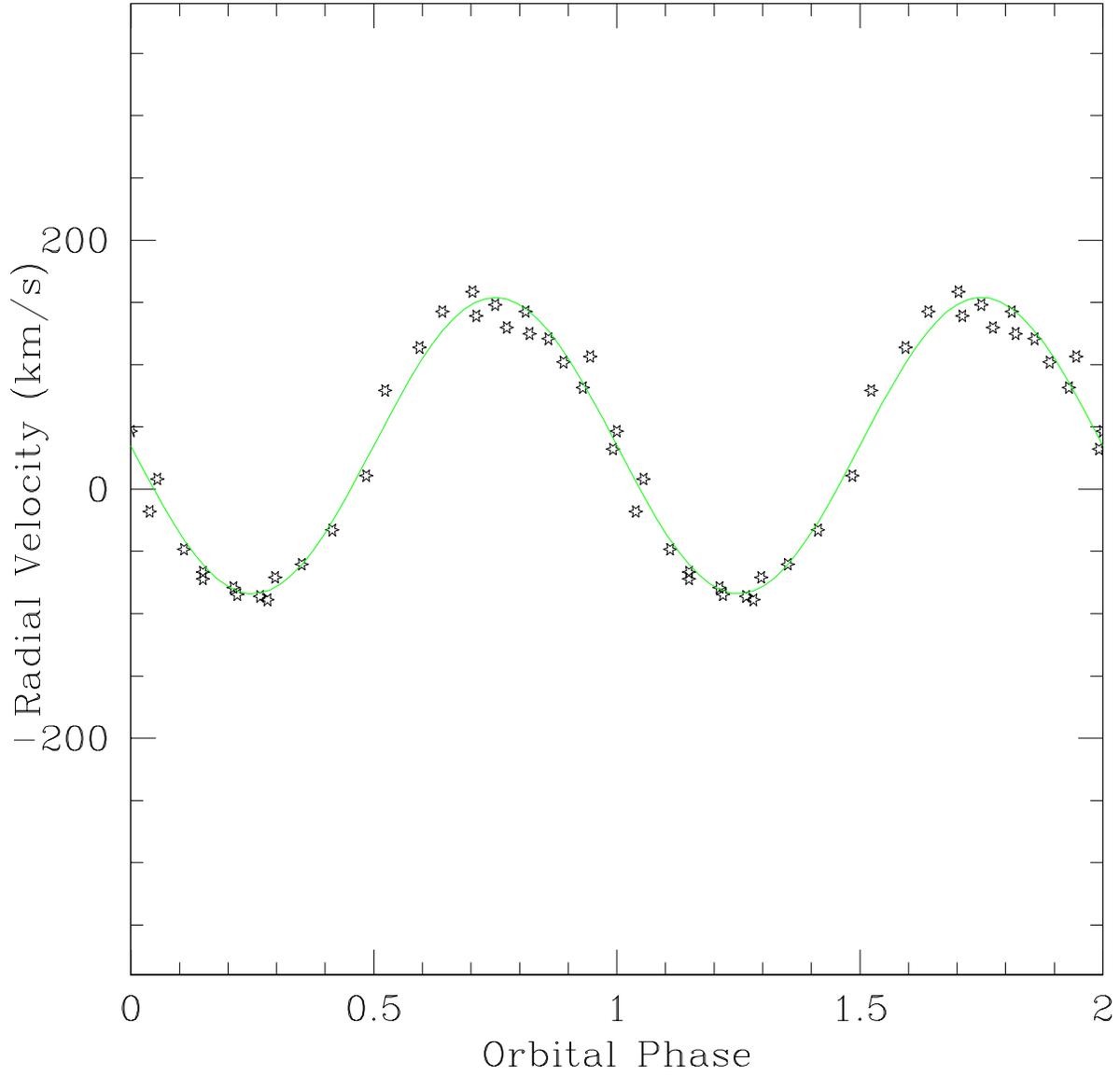}
     \caption{Radial velocities obtained from cross correlation using the template.
     The solid line correspond to the solution from column 3 in Table 2.}
     \label{fig:rvhalfxc}
  \end{center}
\end{figure}

We have also cross-correlated the H$\alpha$ line in our spectra with
a template constructed as follows: First, we selected a spectrum
from the first observed orbital cycle close to phase 0.02 when, in
the case of our observations, we should expect a minimum distortion in the double-peaked
line due to asymmetric components (see section~\ref{tom}). The blue
peak in this spectrum is slightly stronger than the red one. This
is probably caused by the hot spot near the $L_1$ point (see
section~\ref{ugemb}), which might be visible at this phase due to
the fact that the binary has an inclination angle smaller than 70
degrees. The half-separation of the peaks is 470 km
s$^{-1}$, a value similar to that measured in a spectrum taken
during the same orbital phase in next cycle. The chosen spectrum was then
highly smoothed to minimize high-frequency correlations. The
resulting template is shown in Figure \ref{fig:halfxc}.  A radial
velocity for the template was derived from the wavelength measured
at the dip between the two peaks and corrected to give an
heliocentric velocity. The IRAF {\it fxc} task was then used to
derive the radial velocities, which are shown in column 4 of
Table~\ref{tab:RadVel}. As in the previous section, we have fitted
the radial velocities with our nonlinear least-squares fit
algorithm. The resulting orbital parameters are given in column 3 of
Table~\ref{OrbParam}. In Figure~\ref{fig:rvhalfxc} the obtained
velocities and the corresponding sinusoidal fit (solid line) are
plotted.

\begin{figure}[t]
  \begin{center}
     \includegraphics[width=\columnwidth]{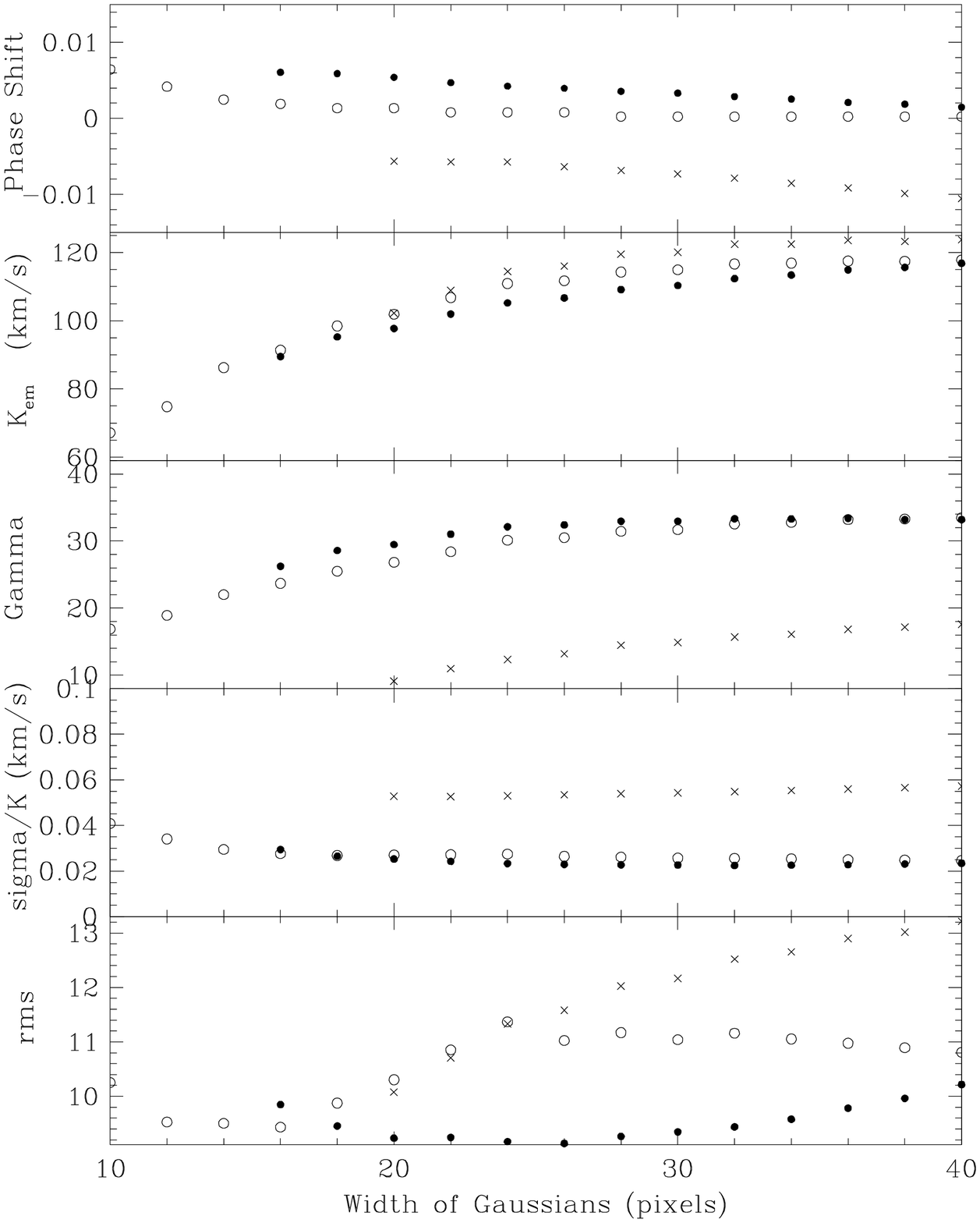}
     \caption{\footnotesize Diagnostic Diagram One. Orbital Parameters as a function of width of individual
     Gaussians for several separations. Crosses correspond to  a = 180 pixels; Dots to a = 230 pixels
     ($\approx 34 \, \rm{\AA}$) and Open circles to a = 280 pixels; }
     \label{fig:diag1}
  \end{center}
\end{figure}

\subsubsection{Measurements from the wings and Diagnostic Diagrams} \label{hdgram}

The H$\alpha$ emission line was additionally measured using the
standard double Gaussian technique and its diagnostic diagrams, as
described in \citet{sha86}. We refer to this paper for the details
on the interpretation of our results. We have used the {\it
convolve} routine from the IRAF {\it rvsao} package, kindly made
available to us by Thorstensen (private communication). The double
peaked H$\alpha$ emission line -- with a separation of about
$20\,\rm{\AA}$ -- shows broad wings reaching up to $40\, \rm{\AA}$
from the line center. Unlike the case of low resolution spectra --
where for over-sampled data the fitting is made with individual
Gaussians having a FWHM of about one resolution element -- in our
spectra, with resolution $\approx$ 0.34 $\rm{\AA}$, such Gaussians
would be inadequately narrow, as they will cover only a very small
region in the wings. To measure the wings appropriately and, at the
same time, avoid possible low velocity asymmetric features, we must
select a $\sigma$ value which fits the line regions corresponding to
disc velocities from about 700 to 1000 $\rm{ km s^{-1}}$.

As a first step, we evaluated the width of the Gaussians by setting
this as a free parameter from 10 to 40 pixels and for a wide range
of Gaussian separations (between 180 and 280 pixels). For each run,
we applied a nonlinear least-squares fit of the computed radial
velocities to sinusoids of the form described in
section~\ref{orbparcal}. The results are shown in
Figure~\ref{fig:diag1}, in particular for three different Gaussian
separations: $a = 180$, 230 and 280 pixels. These correspond to the
low and upper limits as well as to the value for a preferred
solution, all of which are self-consistent with the second step (see
below). In the bottom panel of the figure we have plotted the
overall rms value for each least-squares fit, as this parameter is
very sensitive to the selected Gaussian separations. As expected at
this high spectral resolution, the parameters in the diagram change
rapidly for low values of $\sigma$, and there are even cases when no
solution was found. At low values of $a$ (e.g. crosses) there are no
solutions for widths narrower than 20 pixels. The rms values
increase rapidly with width, while the $\sigma(K)/K$, $\gamma$ and
phase shift values differ strongly from the other cases. For higher
values of $a$ (open circles) we obtain lower values for
$\sigma(K)/K$, but the rms results are still large, in particular
for intermediate values of the width of the Gaussians. For the
middle solution (dots) the results are comparable with those for
large $a$ values, but the rms is much lower. Similar results were
found for other intermediate values of $a$, and they all converge to
a minimum rms for a width of 26 pixels at $a = 230$ pixels.

\begin{figure}[t]
  \begin{center}
     \includegraphics[width=\columnwidth]{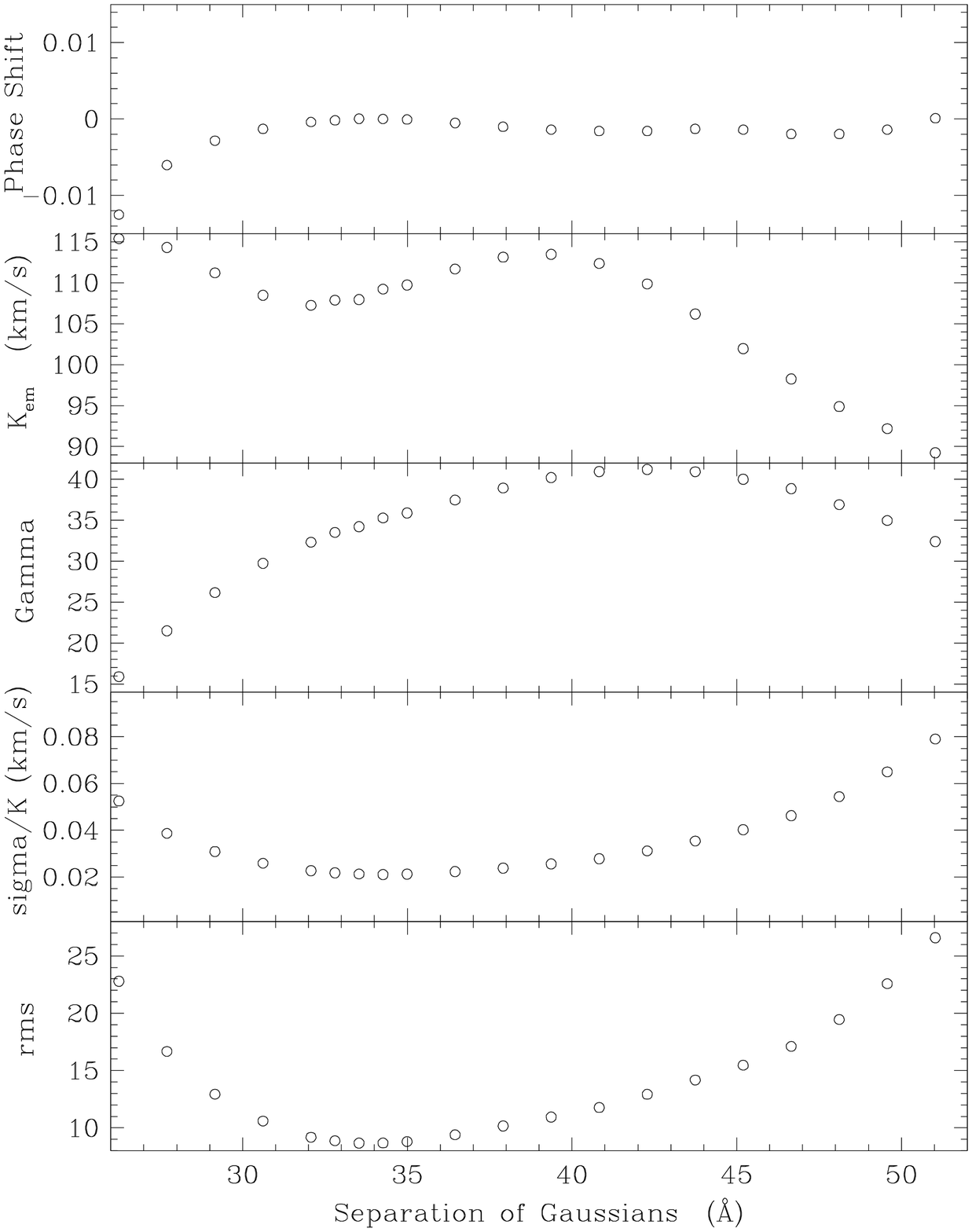}
     \caption{Diagnostic Diagram Two. The best estimate of the semi-amplitude of
     the white dwarf is 107 km s$^{-1}$, corresponding to  a $\approx 34 \, \rm{\AA}$.}
     \label{fig:diag2}
  \end{center}
\end{figure}

For the second step we have fixed the width to a value of 26~$\rm{\AA}$
and ran the double-Gaussian program for a range of $a$ separations,
from about 60 to 120 $\rm{\AA}$. The results obtained are shown in
Figure~\ref{fig:diag2}.

\begin{figure}[t]
  \begin{center}
     \includegraphics[width=\columnwidth]{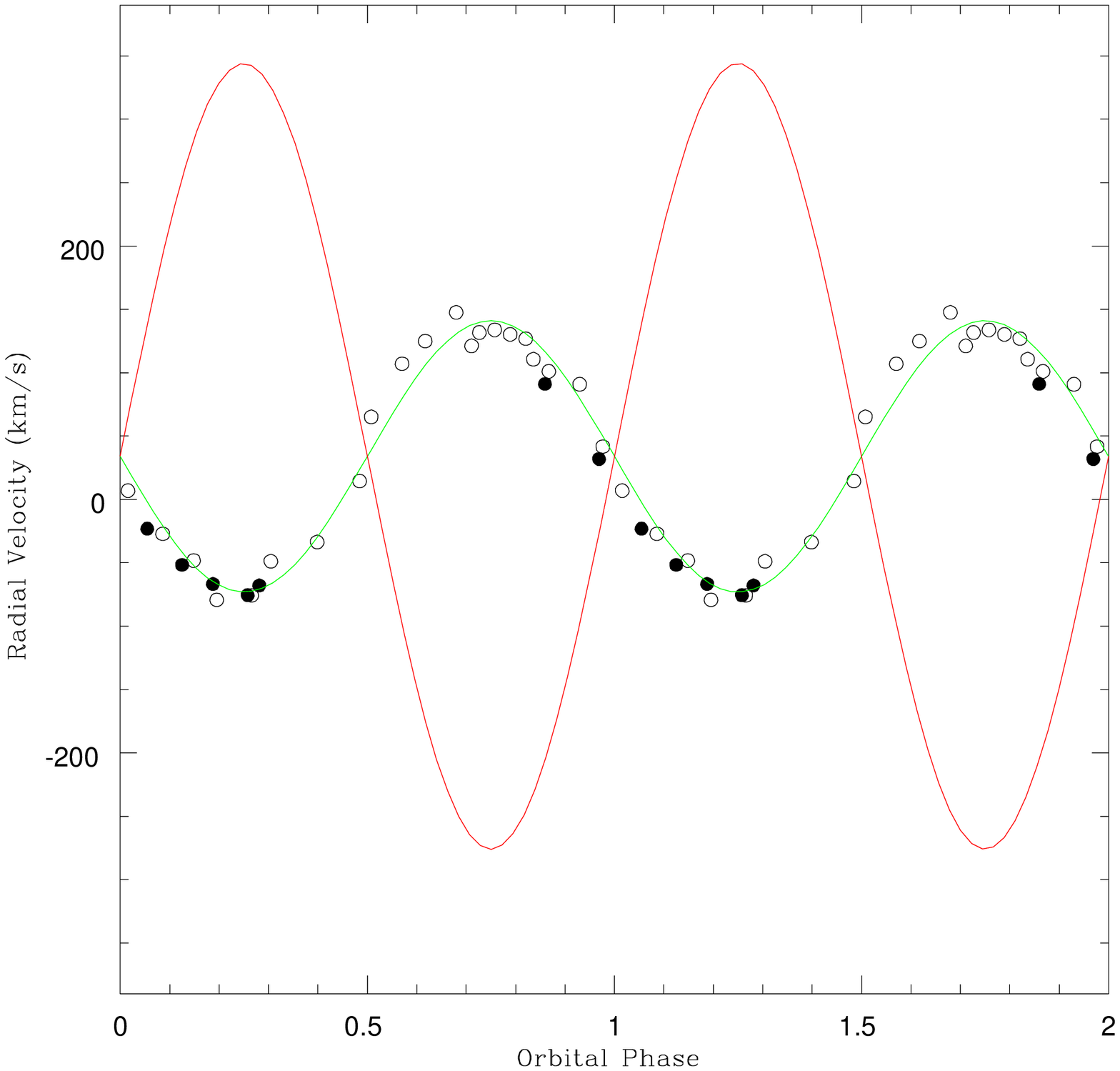}
     \caption{Radial velocities for U Gem. The open circles correspond
     to the measurements of the first 21 spectra single spectra,
     while the dots correspond to those of the co-added spectra
     (see section~\ref{prim}). The solid line, close to the
     points, correspond to the solution with
     $K_{em} = 107 \, {\rm km \, s^{-1}}$ (see text), while the
     large amplitude line correspond to the solution found for
     $K_2$ (see section~\ref{secon}). }
     \label{fig:velrad}
  \end{center}
\end{figure}

If only an asymmetric low velocity component is present, the
semi-amplitude should decrease asymptotically as $a$ increases,
until $K_1$ reaches the correct value. Here we observe such
behavior, although for larger values of $a$, there is a $K_1$
increase for values of $a$ up to  $40 \,\rm{\AA}$, before it
decreases strongly with high values of $a$. This behavior might be
due to the fact that we are observing a narrow hot-spot near the
$L_1$ point (see section~\ref{tom}). On the other hand, as expected,
the $\sigma(K)/K$ vs $a$ curve has a change in slope, at a value of
$a$ for which the individual Gaussians have reached the velocity
width of the line at the continuum. For larger values of $a$ the
velocity measurements become dominated by noise. For low values of
$a$, the phase shift usually gives spurious results, although in our
case it approaches a stable value around 0.015. We believe this
value reflects the difference between the eclipse ephemeris, which
is based mainly on the eclipse of the hot spot, and the true
inferior conjunction of the secondary star. This problem is further
discussed in section~\ref{tom}. Finally, we must point out that the
systemic velocity smoothly increases up to a maximum of about
40~km~s$^{-1}$ at Gaussian separation of nearly 42~\AA, while the
best results, as seen from the Figure, are obtained for $a$~=~31\AA.
This discrepancy may be also be related to the narrow hot-spot near
the $L_1$ point and might be due to the phase-shift between the
hot-spot eclipse and the true inferior conjunction. This problem
will also be address in section~\ref{ephem}. The radial velocities,
corresponding to the adopted solution, are shown in column 5 of
Table~\ref{tab:RadVel} and plotted in Figure~\ref{fig:velrad}, while
the corresponding orbital parameters -- obtained from the nonlinear
least-squares fit -- are given in column 4 of Table~\ref{OrbParam}.

\subsection[]{The Secondary Star} \label{secon}

We were unable to detect single features from the secondary star in
any individual spectra, after careful correction for telluric lines.
In particular we found no radial
velocity results using a standard cross-correlation technique near
the NaI $\lambda\lambda 8183.3, \, 8194.8 \, \mathrm{\AA}$ doublet.
As we will see below, this doublet was very weak compared with
previous observations \citep{wad81, fri90, nay05}. We have been
able, however, to detect the NaI doublet and the TiO Head band
around $\lambda$7050~\AA~with a new technique, which enables us to
derive the semi-amplitude $K_{abs}$ of the secondary star velocity
curve. We first present here the general method for deriving the
semi-amplitude and then apply it to U Gem, using not only the
absorption features but the $H\alpha$ emission as well.

\subsubsection{A new method to determine K$_2$} \label{secon}

In many cataclysmic variables the secondary star is poorly visible, or even absent,
in the optical spectral range. Consequently, no $V(t)$
measurements are feasible for this component. Among these systems
are dwarf novae with orbital periods under 0.25 days, for which it
is thought that the disc luminosity dominates over the luminosity of
the Roche-Lobe filling secondary, whose brightness depends on the
orbital period of the binary \citep{ech84}. For such binaries, the
orbital parameters have been derived only for the white-dwarf-
accretion disc system, in a way similar to that described in
section~\ref{orbparcal}.

In order to determine a value of $K_{abs}$ from a set of spectra of
a cataclysmic variable, for which the orbital period and time of
inferior conjunction have been already determined from the emission
lines, we propose to reverse the process: derive $V(t)_{abs}$ using
$K_{pr}$ as the initial value for the semi-amplitude, and set the
values of $P_{orb}$ and $\mathrm{HJD_{\odot}}$, derived from the
emission lines, as constants. The initial value for the systemic
velocity is set to zero, and its final value may be calculated later
(see below). The individual spectra are then {\it co-added} in the
frame of reference of the secondary star, i.e. by Doppler-shifting
the spectra using the calculated $V(t)_{calc}$ from the equation
given in section~\ref{orbparcal}, and then add them together.
Hereinafter we will refer to this procedure as the {\it co-phasing
process}. Ideally, as the proposed $K_{pr}$ is changed through a
range of possible values, there will be a one for which the {\it
co-phased} spectral features associated with the absorption spectrum
will have an optimal signal-to-noise ratio.

In fact, this will also be the case for any emission line features
associated with the red star, if present. In a way, this process
works in a similar fashion as the double Gaussian fitting used in
the previous section, provided that adequate criteria are set in
order to select the best value for $K_{abs}$. We propose three
criteria or tests that, for late type stars, may be used with this
method: The first one consists in analyzing the behavior of the
measured depths or widths of a well identified absorption line in
the {\it co-phased} spectra, as a function of the proposed $K_{pr}$;
one would expect that the width of the line will show a minimum and
its depth a maximum value at the optimal solution. This method could
be particularly useful for K-type stars which have strong single
metallic lines like Ca I and Fe I. The second criterion is based
upon measurements of the slope of head-bands, like that of TiO at
$\lambda$7050~\AA. It should be relevant to short period systems,
with low mass M-type secondaries with spectra featuring strong
molecular bands. In this case one could expect that the slope of the
head-band will be a function of $K_{pr}$, and will have a maximum
negative value at the best solution. A third test is to measure the
strength of a narrow emission arising from the secondary. This
emission, if present, would be particularly visible in the {\it
co-phased} spectrum and will have minimum width and maximum height
at the best selected semi-amplitude $K_{pr}$.

We have tested these three methods by means of an artificial spectrum
with simulated narrow absorption lines, a TiO-like head band and a
narrow emission line. The spectrum with these artificial features
was then Doppler shifted using pre-established inferior
conjunction phase and orbital period, to produce a series of test
spectra. An amount of random Gaussian noise was added to each
Doppler shifted spectrum, sufficient to mask the artificial
features. We then proceeded to apply the {\it co-phasing process} to
recover our pre-determined orbital values. All three criteria
reproduced back the original set of values, as long as the random
noise amplitude was of the same order of magnitude as the strength
of the {\it clean} artificial features.

\subsubsection{Determination of K$_2$ for U Gem} \label{ugemb}

We have applied the above-mentioned criteria to U Gem. The time of
the inferior conjunction of the secondary and the orbital period
were taken from section~\ref{ephem}. To attain the best signal to
noise ratio we have used all the 28 observed spectra. Although they
span over slightly more than 1.5 orbital periods, any departure from
a real $K_2$ value will not depend on selecting data in exact
multiples of the orbital period, as any possible deviation from the
real semi-amplitude will already be present in one complete orbital
period and will depend mainly on the intrinsic intensity
distribution of the selected feature around the secondary itself
(also see below the results for $\gamma$).

\begin{figure}[!]
  \begin{center}
     \includegraphics[width=\columnwidth]{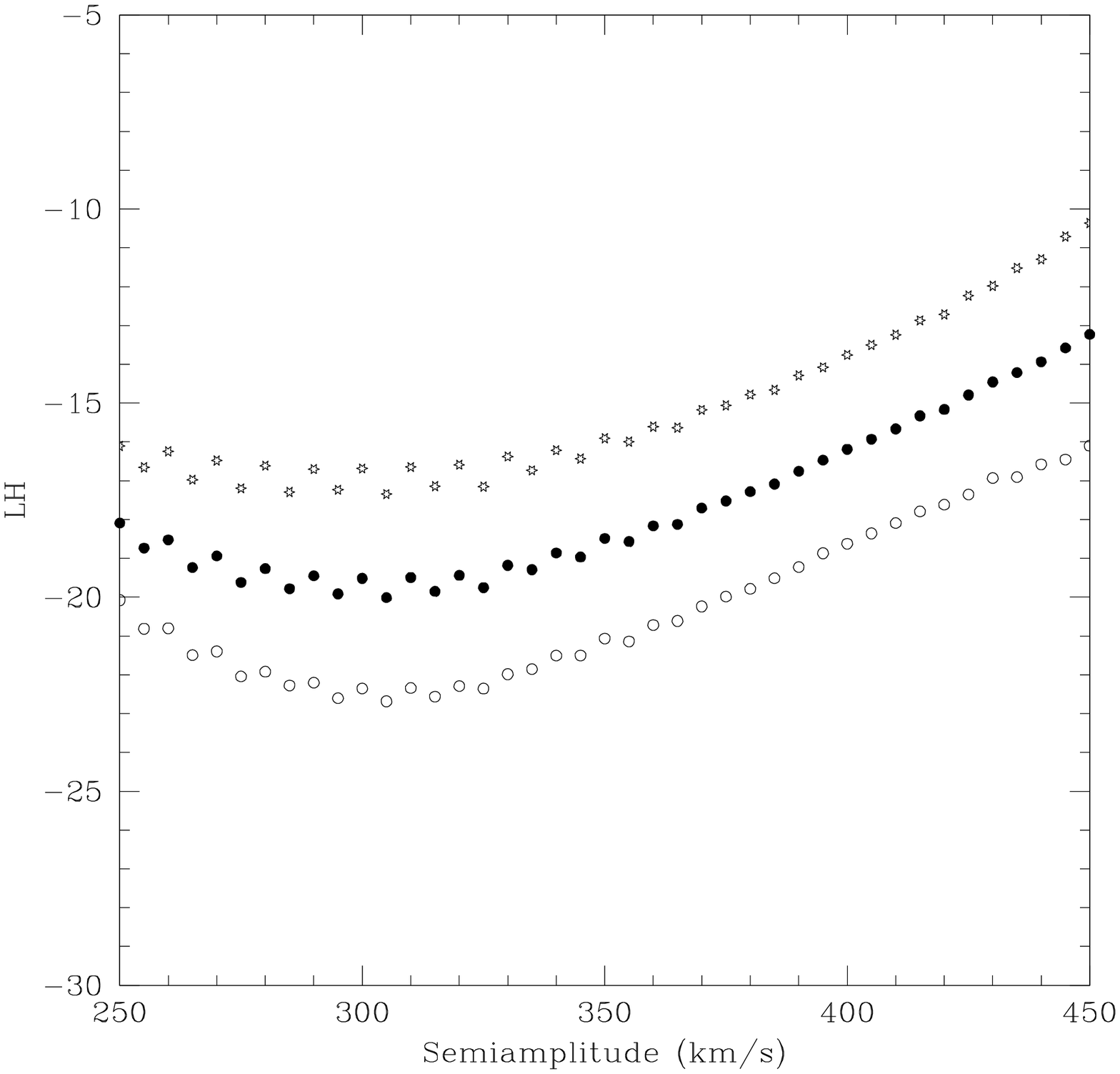}
     \caption{Maximum flux depth of the individual NaI lines $\lambda 8183.3 \, \mathrm{\AA} $ (top),
     $\lambda 8194.8 \, \mathrm{\AA}$ (bottom) and mean (middle) as a function of $K_{pr}$.}
     \label{fig:sodcor}
  \end{center}
\end{figure}

\begin{figure}[!]
  \begin{center}
     \includegraphics[width=0.9\columnwidth,angle=90]{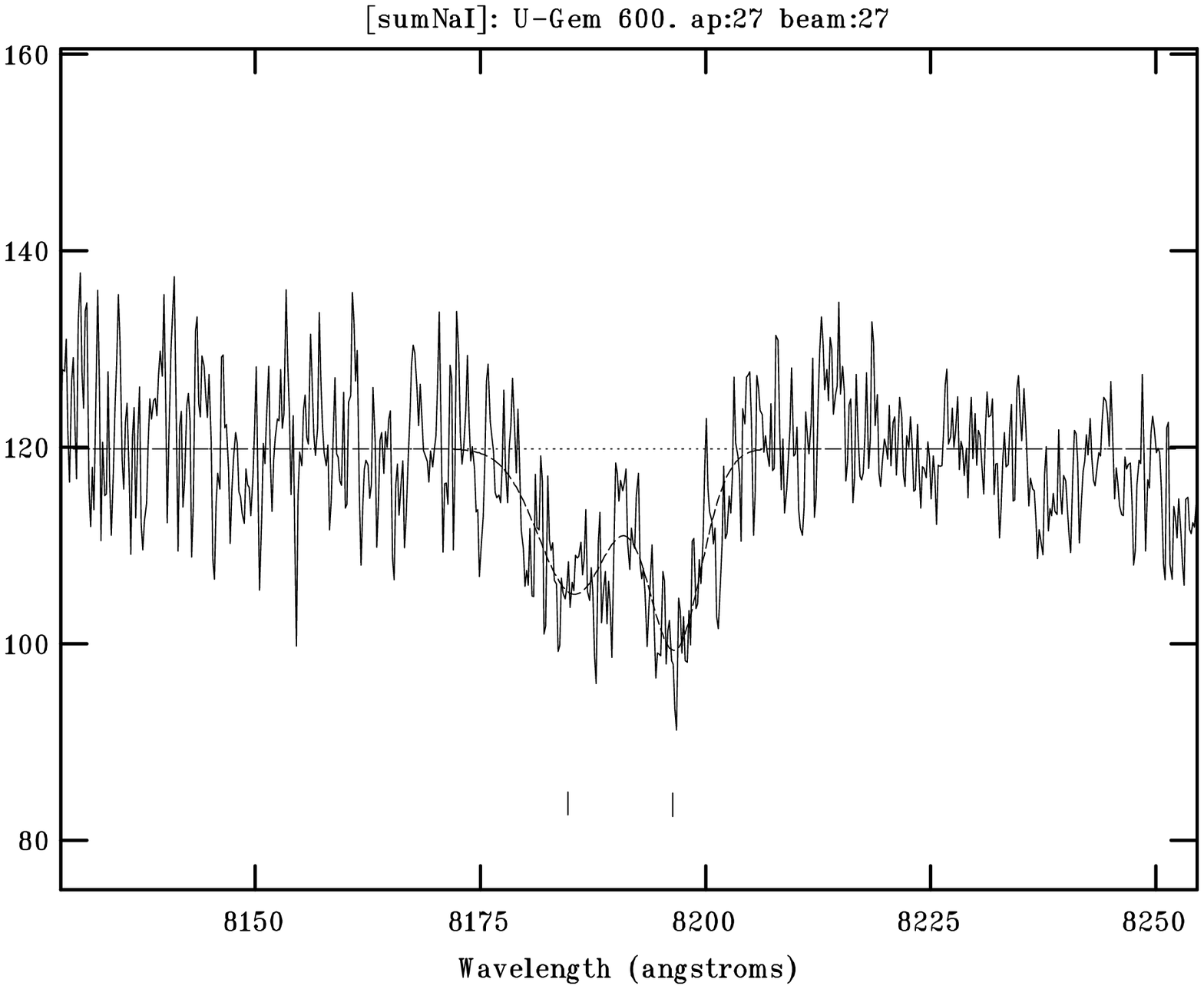}
     \caption{Co-phased spectrum around the NaI doublet.}
     \label{fig:sodspec}
  \end{center}
\end{figure}

Figure \ref{fig:sodcor} shows the  application of the first test
to the NaI doublet $\lambda\lambda$~8183,8195~\AA. The spectra were
{\it co-phased} varying $K_{pr}$ between 250 to 450 km~s$^{-1}$. The
line depth of the blue and red components of the doublet (stars and
open circles, respectively), as well as their mean value (dots) are
shown in the diagram. We find a best solution for
$K_2$~=~310~$\pm$~5~km~s$^{-1}$. The error has been estimated from
the intrinsic modulation of the solution curve. As it approaches its
maximum value, the line depth value oscillates slightly, but in the
same way for both lines. A similar behavior was present when low
signal to noise features were used on the artificial spectra process
described above. Figure~\ref{fig:sodspec} shows the {\it co-phased}
spectrum of the NaI doublet of our best solution for $K_2$. These
lines appear very weak as compared with those reported by
\citet{fri90} and \citet{nay05}. We have also measured the gamma
velocity from the co-phased spectrum by fitting a double-gaussian to
the Na~I doublet (dotted line in Figure~\ref{fig:sodcor}) and find a
mean value $\gamma = 69 \,\pm$ 10 km s$^{-1}$ (corrected to the
heliocentric standard of motion). We did a similar calculation for
$\gamma$ by {\it co-phasing} the selected spectra used in
section~\ref{tom}, covering a full cycle only. The results were very
similar to those obtained by using all spectra.

The second test, to measure the slope of the TiO band has at
$\lambda$7050~\AA~was not successful. The solution curve oscillates
strongly near values between 250 and 350 ${\rm km \, s^{-1}}$. We
believe that the signal to noise ratio in our spectra is too poor for this test and
that more observations, accumulated during several orbital cycles,
have to be obtained in order to attain a reliable result using this
method.

\begin{figure}[!]
  \begin{center}
     \includegraphics[width=1.2\columnwidth]{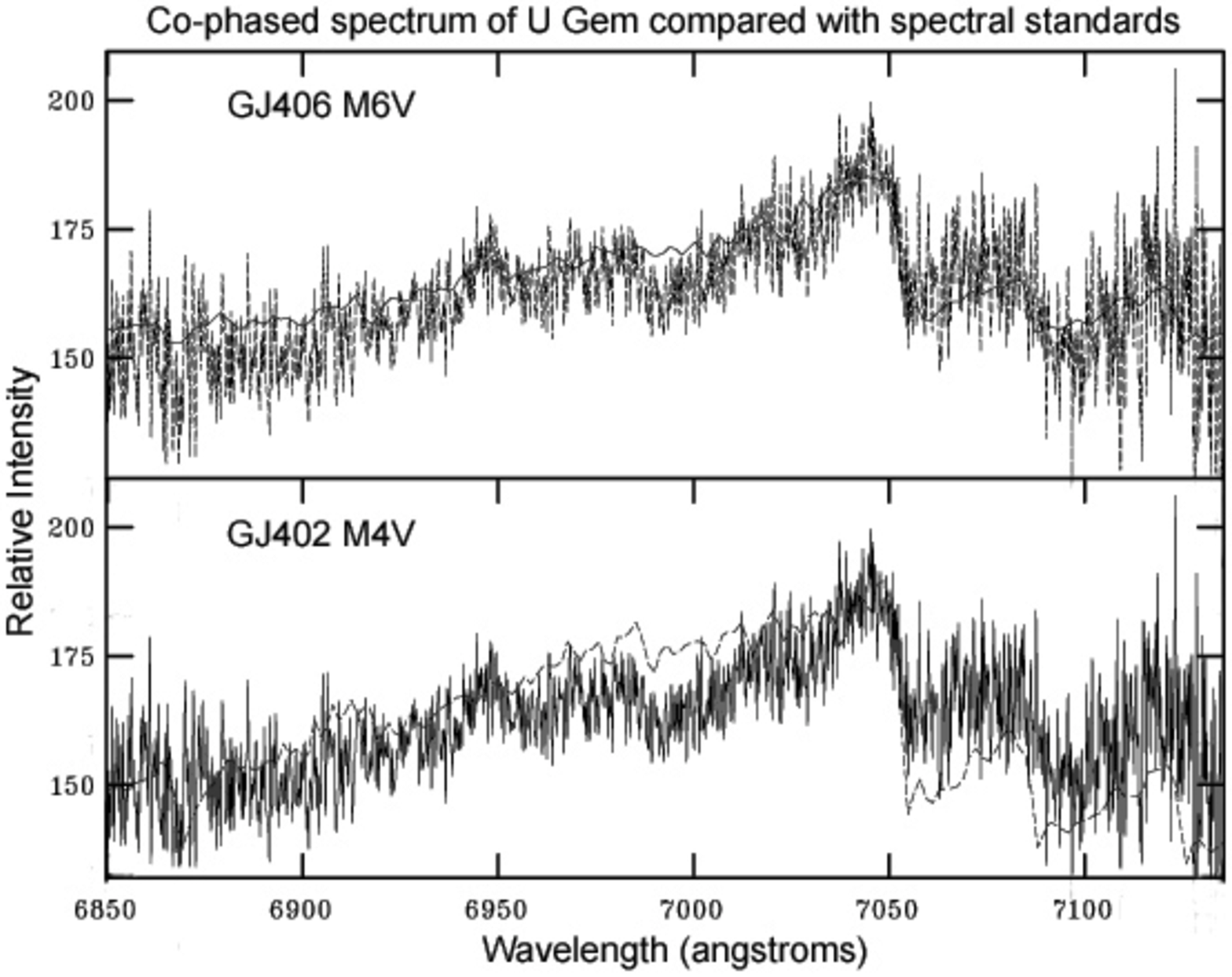}
     \caption{U Gem TiO Head Band near 7050~\AA~compared with GJ406, an M6V star
     (upper diagram), and GJ402, an M4~V star (lower diagram) (see text).}
     \label{fig:tioband}
  \end{center}
\end{figure}

However, we have co-phased our spectra for $K_2$~=~310 km~s$^{-1}$, with
the results shown in Figure~\ref{fig:tioband}. The TiO band is clearly
seen while the noise is prominent,
particulary along the slope of the head-band. We have used this
co-added spectrum to compare it with several late-type M stars
extracted from the published data by \citet{mon97}
fitted to our {\it co-phased} spectrum. A gray continuum
has been added to the comparison spectra in order to compensate for
the fill-in effect arising from the other light sources in the
system, so as to obtain the best fit. In particular, we show in the same figure the
fits when two close candidates -- GJ406 (M6~V, upper panel) and GJ402
(M4-5~V, lower panel) -- are used. The best fit is obtained for the
M6~V star, to which we have added a 95 percent continuum. For the
M4-5~V star the fit is poor, as we observe a flux excess around
7000 \AA~and a stronger TiO head-band. Increasing the grey flux
contribution will fit the TiO head band, but will result in a larger
excess at the 7000 \AA~region. On the other hand, the fit with the
M6~V star is much better all along the spectral interval. There are
a number of publications which assign to U~Gem spectral types M4
\citep{har00}, M5 \citep{wad81} and possibly as far as M5.5
\citep{ber83}. Even in the case that the spectral type of the
secondary star were variable, its spectral classification is still
incompatible with its mass determination \citep{ech83}.

\begin{figure}[!]
  \begin{center}
     \includegraphics[width=\columnwidth]{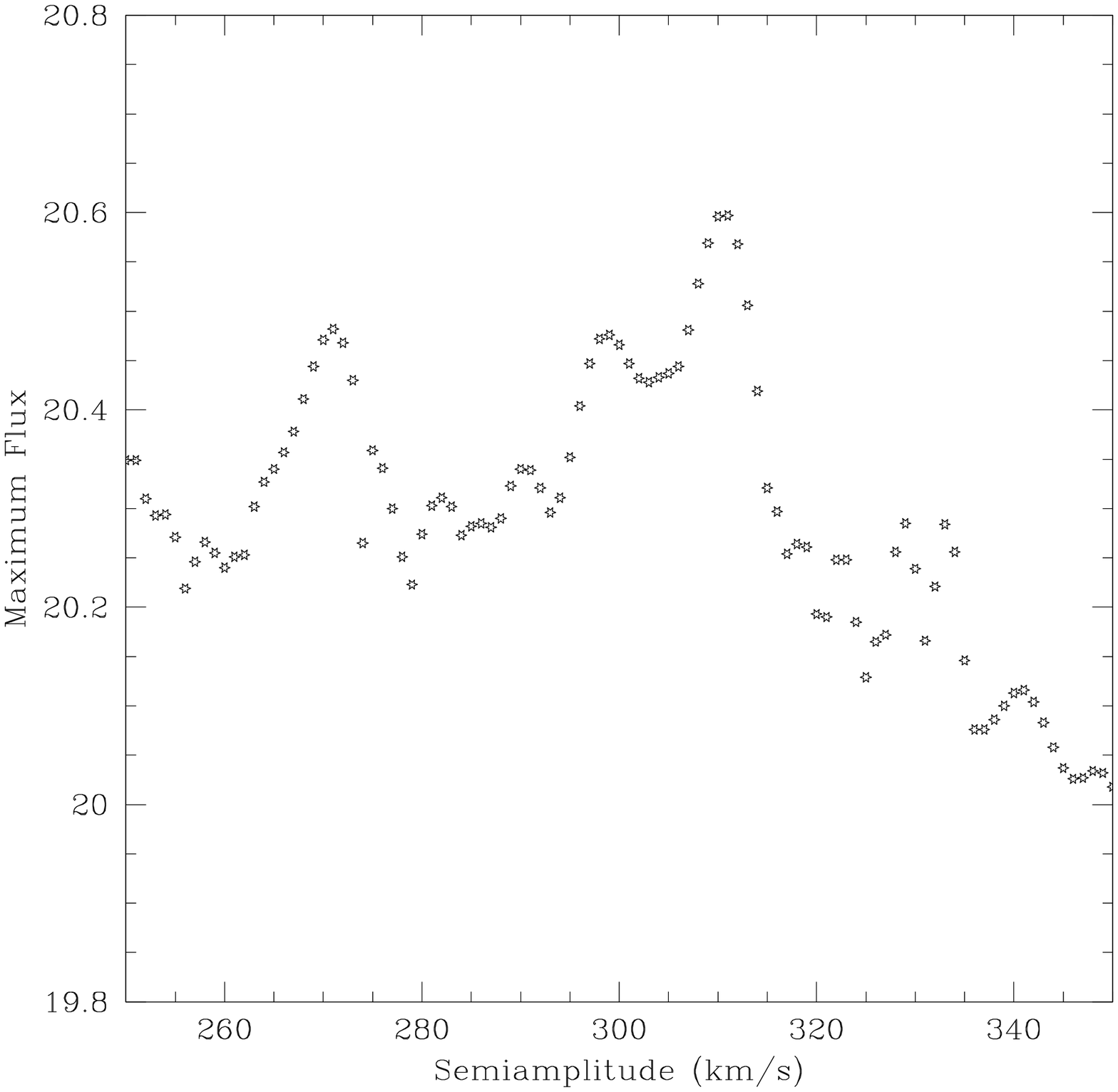}
     \caption{Maximum peak flux of the co-added $H\alpha$ spectra as a function of $K_{pr}$ }
     \label{fig:hotflux}
  \end{center}
\end{figure}

\begin{figure}[!]
  \begin{center}
     \includegraphics[width=0.8\columnwidth,angle=90]{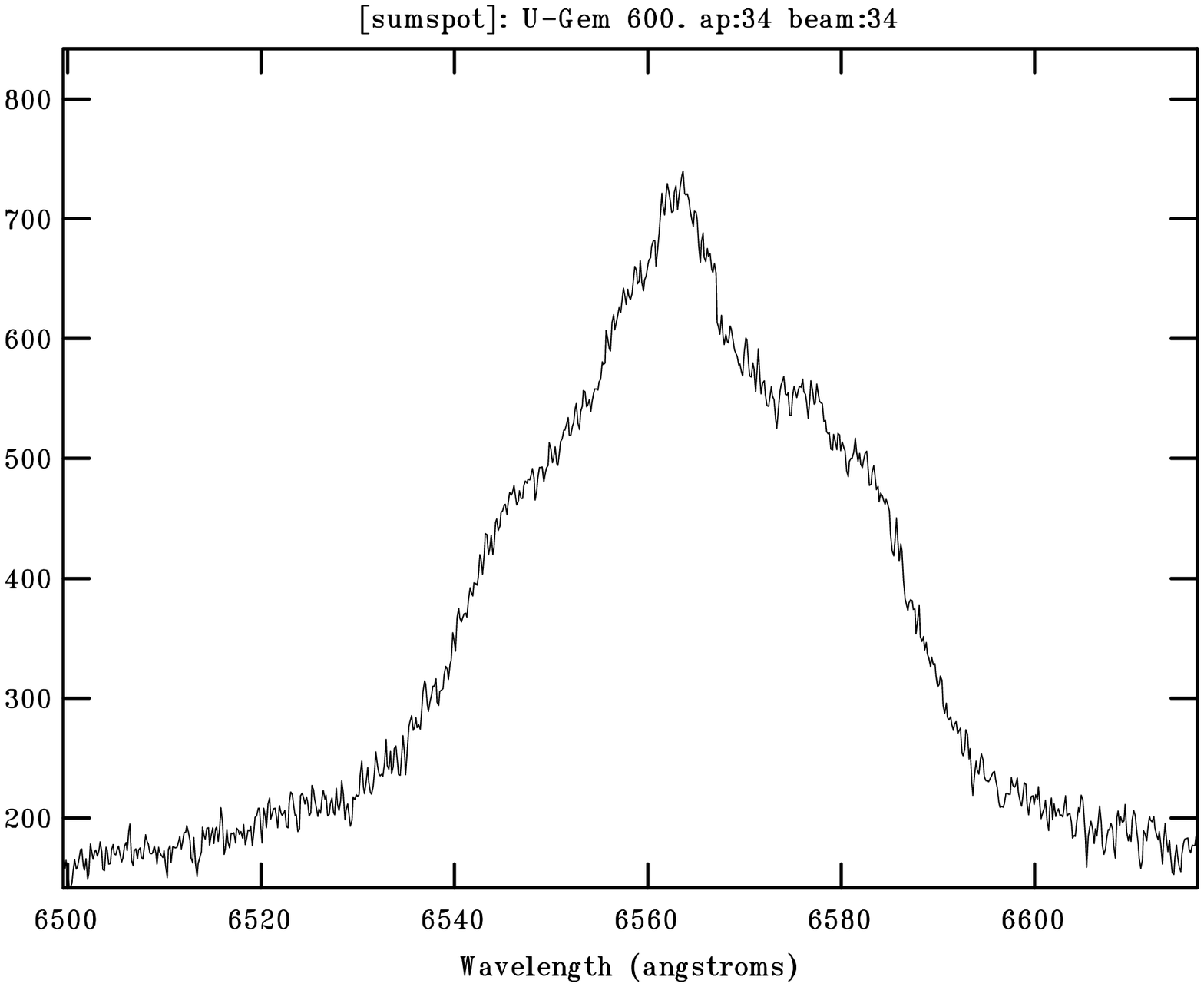}
     \caption{Shape of the co-added $H\alpha$ spectrum for
     $\mathrm{K_{2}} = 310 \,{\rm km \, s^{-1}}$.}
     \label{fig:maxiflux}
  \end{center}
\end{figure}

For the third test, we have selected the region around H$\alpha$, as
in the individual spectra we see evidence of a narrow spot, which is
very well defined in our spectrum near orbital phase 0.5. In this
test we have {\it co-phased} the spectra as before, and have adopted
as the test parameter the  peak intensity around the emission line.
The results are shown in Figure~\ref{fig:hotflux}. A clear and
smooth maximum is obtained for $K_{pr}$~=~310~$\pm$~3~km~s$^{-1}$.
The co-phased spectrum obtained from this solution is shown in
Figure~\ref{fig:maxiflux}. The double-peak structure has been
completely smeared -- as expected when co-adding in the reference
frame of the secondary star, as opposed to that of the primary star-
and instead we observe a narrow and strong peak at the center of the
line. We have also fitted the peak to find the radial velocity of
the spot. We find $\gamma = 33 \,\pm$ 10 km~s$^{-1}$, compatible
with the gamma velocity derived from the radial velocity analysis of
the emission line, $\gamma = 34 \,\pm$ 2 km~s$^{-1}$ (see
section~\ref{hdgram}). This is a key result for the determination of
the true systemic velocity and can be compared with the values
derived from the secondary star (see section~\ref{discus}).

\section{Improved Ephemeris of U Gem} \label{ephem}

As mentioned in section~\ref{orbparcal}, the presence of eclipses in
U Gem and an ample photometric coverage during 30 years has
permitted to establish, with a high degree of accuracy, the value of
orbital period. This has been discussed in detail by \citet{mar90}.
However, as pointed by these authors, this object shows erratic
variations in the timing of the photometric mid-eclipse that may be
caused either by orbital period changes, variations in the position
of the hot spot, or they may even be the consequence of the
different methods of measuring of the eclipse phases. A variation in
position and intensity of the gas stream will also contribute to
such changes. A date for the zero phase determined independently
from spectroscopic measurements would evidently be desirable.
\citet{mar90} discuss two spectroscopic measurements by
\citet{mar88} and \citet{wad81}, and conclude that the spectroscopic
inferior conjunction of the secondary star occurs about 0.016 in
phase prior to the mean photometric zero phase. There are two
published spectroscopic studies \citep{hon87, sto81}, as well as one
in this paper, that could be used to confirm this result.
Unfortunately there is no radial velocity analysis in the former
paper, nor in the excellent Doppler Imaging paper by \citet{mar90}
based on their original observations. However, the results by
\citet{sto81} are of particular interest since he finds the
spectroscopic conjunction in agreement with the time of the eclipse
when using the photometric ephemerides by \citet{wad81}, taken from
\citet{arn76}. The latter authors introduce a small quadratic term
which is consistent with the O-C oscillations shown in
\citet{mar90}.

It is difficult to compare results derived from emission lines to
those obtained from absorption lines, especially if they are based
on different ephemerides. Furthermore, the contamination on the
timing of the spectroscopic conjunction -- either caused by a hot
spot, by gas stream or by irradiation on the secondary -- has not
been properly evaluated. However, since our observations were made
at a time when the hot spot in absent (or, at least, is along the
line between the two components in the binary) and the disc was very
symmetric (see section~\ref{tom}), we can safely assume that in our
case, the photometric and spectroscopic phases must coincide. If we
then take the orbital period derived by \citet{mar90} and use the
zero point value derived from our measurements of the H$\alpha$
wings, (section~\ref{hdgram}), we can improve the ephemeris:

$$\rm{HJD} = 2,437,638.82566(4) \, + \, 0.1769061911(28) \,E \, ,$$

for the inferior conjunction of the secondary star. These ephemeris
are used throughout this paper for all our phase folded diagrams and
Doppler Tomography.

\begin{figure}[h]
\centerline{
  \hbox{
    \resizebox{45mm}{!}
    {\includegraphics{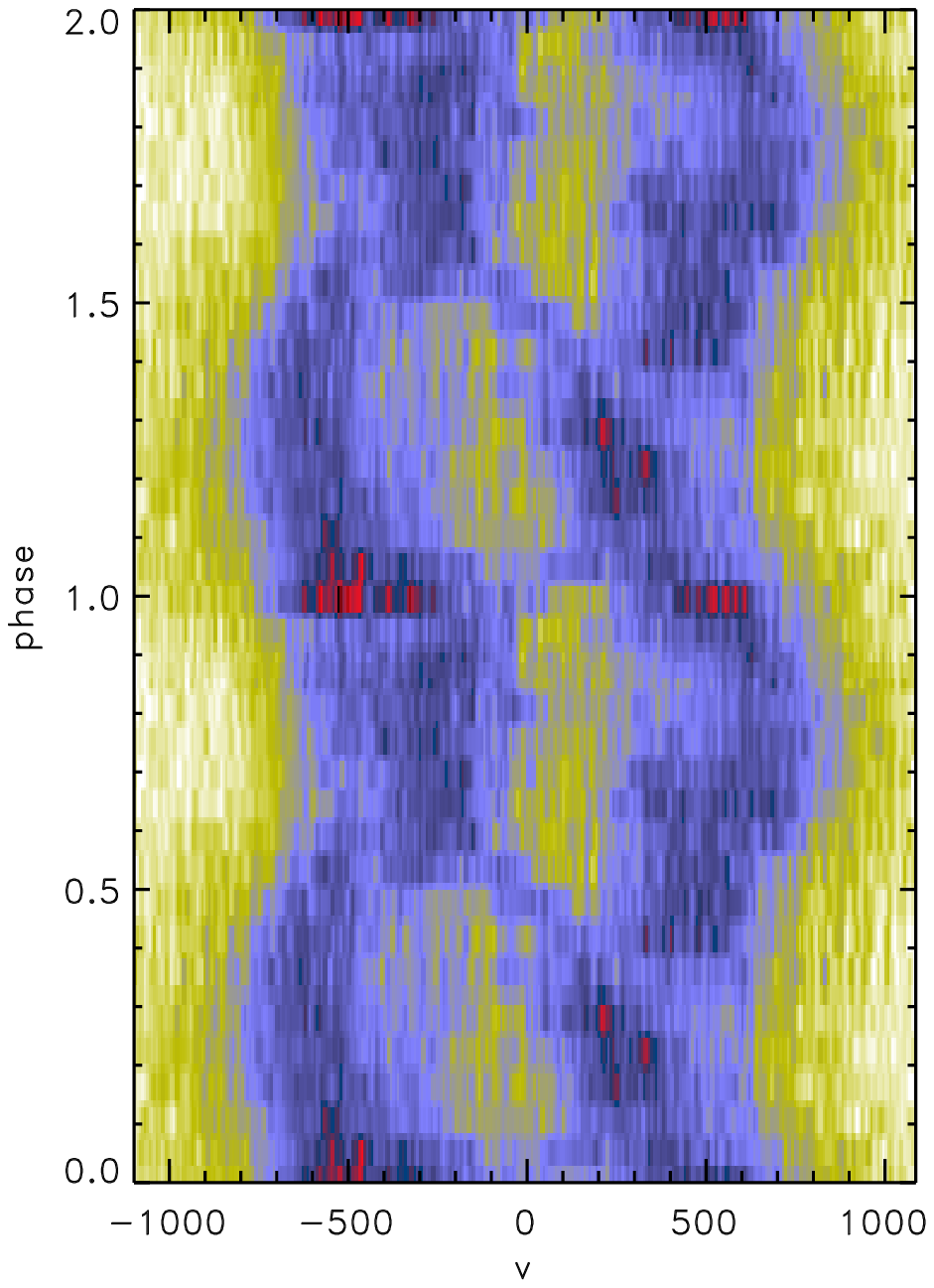}}
    \resizebox{45mm}{!}
    {\includegraphics{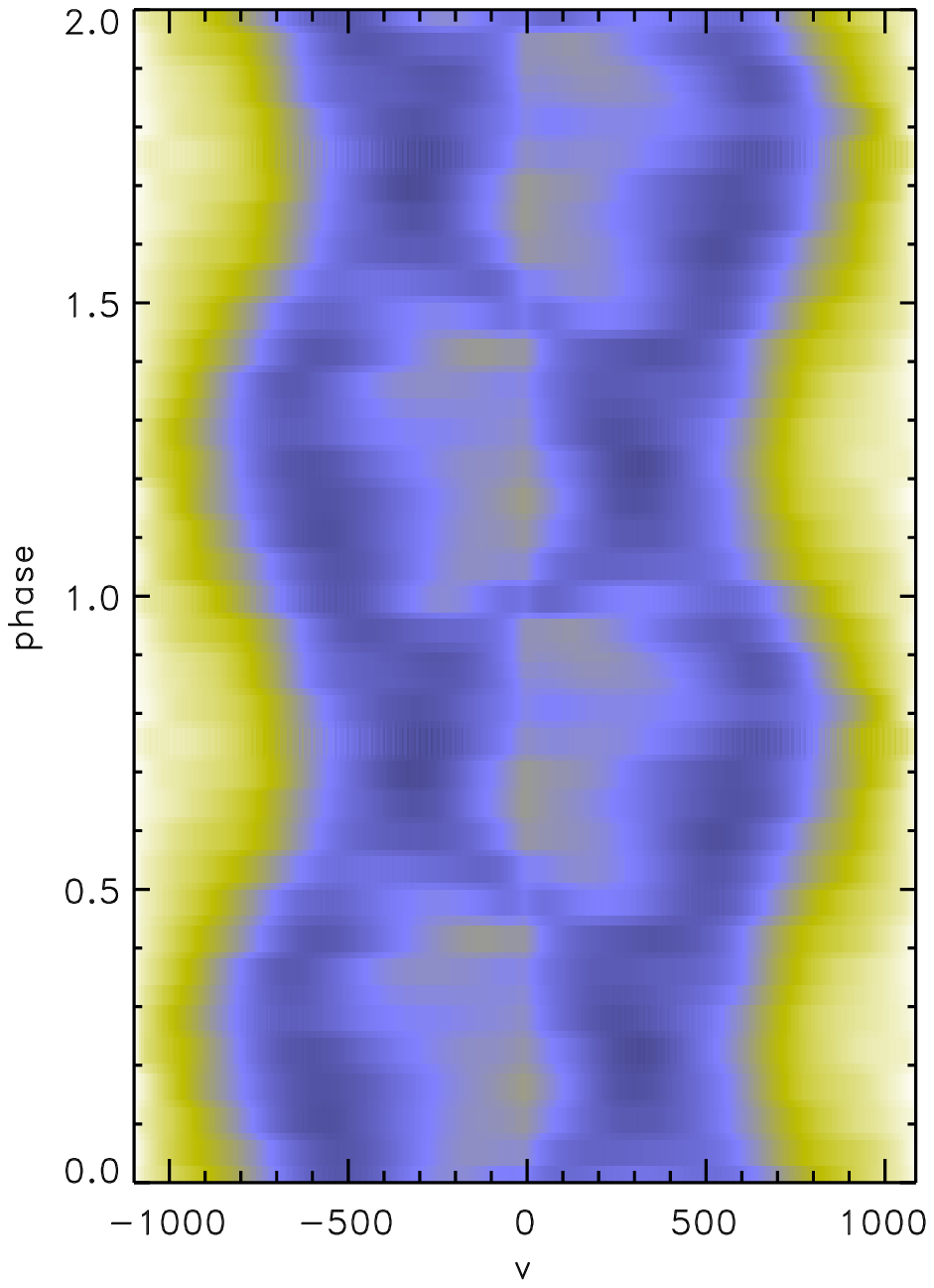}}
  }
}
\caption{Trailed spectra of the H$\alpha$ emission line. Original (left) and
         reconstructed data (right).}
\label{fig:spec2}
\end{figure}

\section{Doppler Tomography} \label{tom}

Doppler Tomography is a useful and powerful tool to study the
material orbiting the white dwarf, including the gas stream coming
from the secondary star as well as emission regions arising from the
companion itself. It uses the emission line profiles observed as a
function of the orbital phase to reconstruct a two-dimensional
velocity map of the emitting material. A detailed formulation of
this technique can be found in \citet{mar88}. A careful
interpretation of these velocity maps has to be made, as the main
assumption invoked by tomography is that all the observed material
is in the orbital plane and is visible at all times.

\begin{figure*}
\epsscale{.9} \plotone{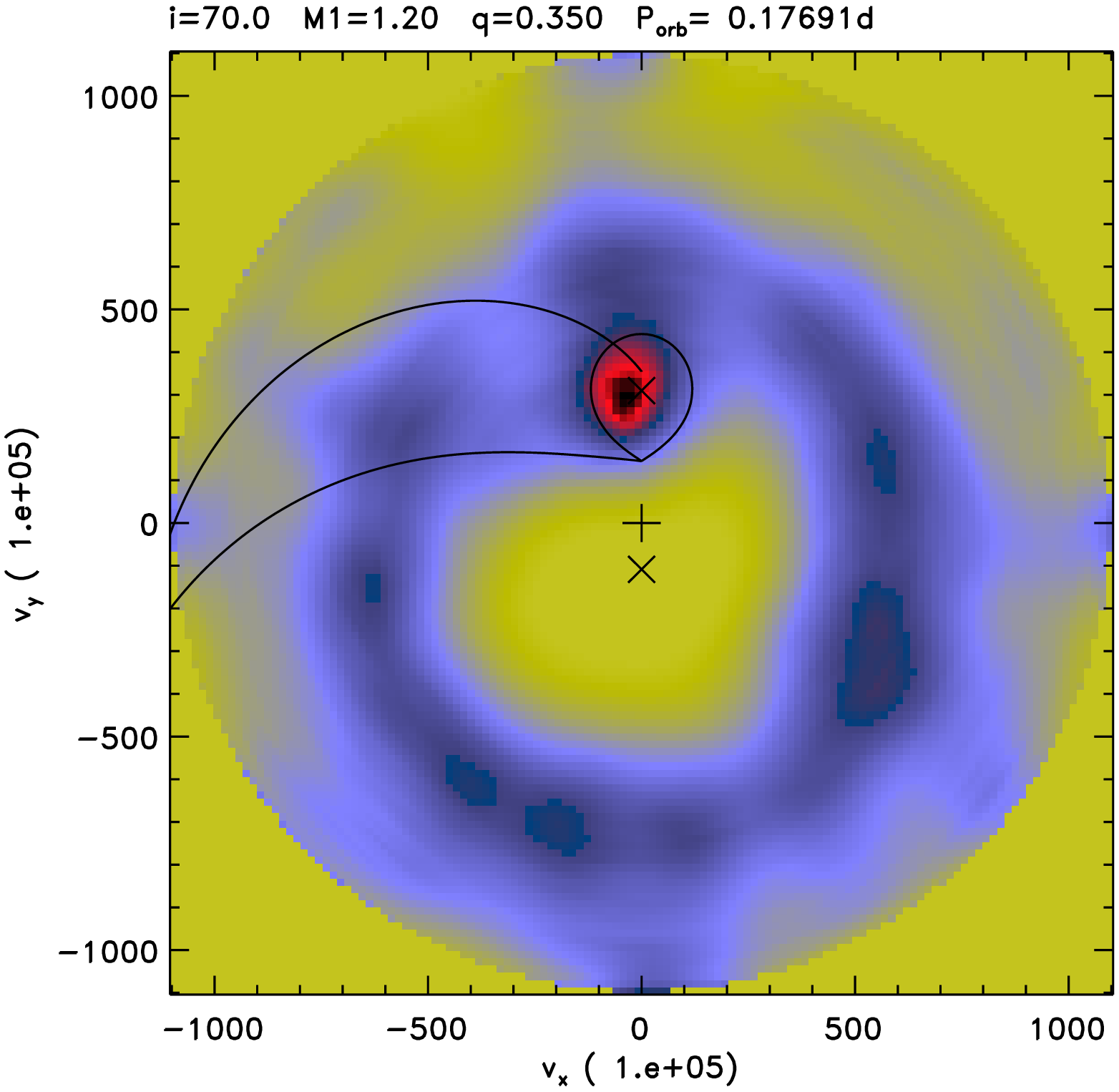}
  \caption{Doppler Tomography of U~Gem. The various features are discussed in the text. The $v_x$ and $v_y$ axes are in
  km s$^{-1}$. A compact hot spot, close to the inner Lagrangian point is detected instead of the usual bright spot and/or broad stream,
  where the material, following a Keplerian or ballistic trajectory strikes the disc. The Tomogram reveals
  a full disc whose outer edge is very close to the L$_1$ point (see text).}
  \label{fig:doptom}
\end{figure*}

\begin{figure*}
\epsscale{1} \plotone{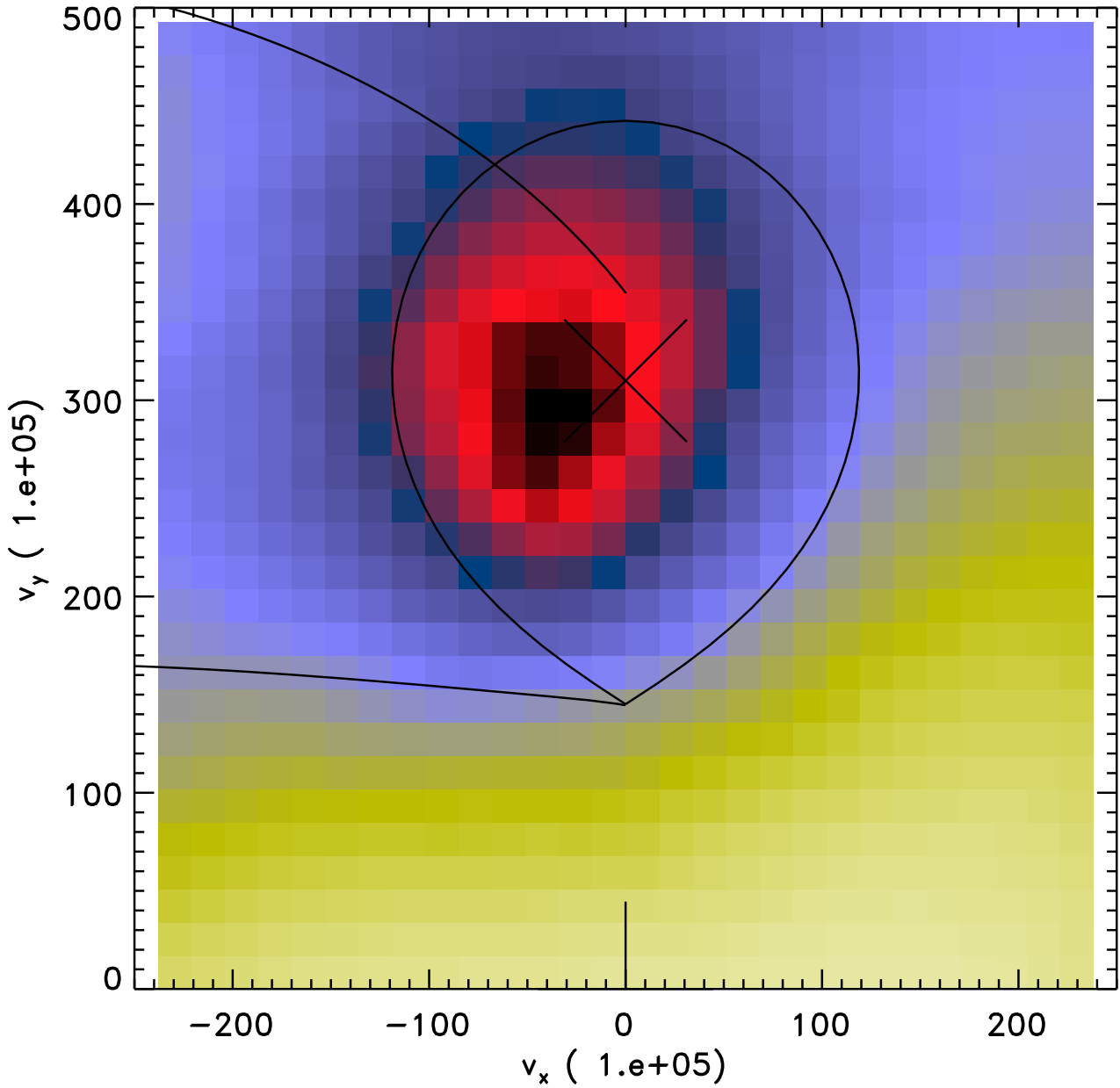}
  \vspace{20mm}
  \caption{Blow-up of the region around the hot spot. Note that this feature is slightly ahead of the center of mass of
  the secondary star. Since this is a velocity map and not a geometrical one, its physical position in the binary
  is carefully discussed in the text.}
  \label{fig:spottom}
\end{figure*}

The Doppler Tomography, derived here from the H$\alpha$ emission
line in U~Gem, was constructed using the code developed by
\citet{spr98}. Our observations of the object cover 1.5 orbital
cycles. Consequently -- to avoid disparities on the intensity of the
trailed and reconstructed spectra, as well as on the tomographic map
-- we have carefully selected spectra covering a full cycle only.
For this purpose we discarded the first 3 spectra (which have the
largest airmass) and used only 18 spectra out of the first 21, 600~s
exposures, starting with the spectrum at orbital phase 0.88 and
ending with the one at phase 0.86 (see Table~\ref{tab:RadVel}). In
addition, in generating the Tomography map we have excluded the
spectra taken during the partial eclipse of the accretion disc
(phases between 0.95 and 0.05). The original and reconstructed
trailed spectra are shown in Figure~\ref{fig:spec2}. They show the
sinusoidal variation of the blue and read peaks, which are strong at
all phases. The typical S-wave is also seen showing the same simple
sinusoidal variation, but shifted by 0.5 in orbital phase with
respect to the double-peaks. The Doppler tomogram is shown in
Figure~\ref{fig:doptom}; as customary, the oval represents the
Roche-Lobe of the secondary and the solid lines the Keplerian
(upper) and ballistic (lower) trajectories. The Tomogram reveals a
disc reaching to the distance of to the inner Lagrangian point in
most phases. A compact and strong emission is seen close to the
center of velocities of the secondary star. A blow-up of this region
is shown in Figure~\ref{fig:spottom}. Both maps have been constructed
using the parameters shown at the top of the diagrams and a $\gamma$
velocity of 34 km~s$^{-1}$. The velocity resolution of the map near
the secondary star is about 10 km s$^{-1}$. The $V(x,y)$ position of
the hot-spot (in km~s$^{-1}$) is (-50,305), within the
uncertainties.

The tomography shown in Figure~\ref{fig:doptom} is very different
from what we expected to find and from what has been observed by
other authors. We find a very symmetric full disc, reaching close to
the inner Lagrangian point and a compact bright spot also close to
the L$_1$ point, instead of a complex system like that observed by
\citet{und06}, who find U~Gem at a stage when the Doppler Tomographs
show: emission at low velocity close to the center of mass; a
transient narrow absorption in the Balmer lines; as well as two
distinct spots, one very narrow and close in velocity to the
accretion disc near the impact region and another much broader,
located between the ballistic and Keplerian trajectories. They
present also tentative evidence of a weak spiral structure, which
have been seen as strong spiral shocks during an outburst observed
by \citet{gro01}. Our results also differ from those of
\citet{mar90}, who also find that the bulk of the bright spot
arising from the Balmer, He~I and He~II emission come from a region
between the ballistic and Keplerian trajectories. We interpret the
difference between our results and previous studies simply by the
fact that we have observed the system at a peculiar low state not
detected before (see sections~\ref{intro} and \ref{discus}) . This
should not be at all surprising because, although U~Gem is a well
observed object, it is also a very unusual and variable system.

Figure~\ref{fig:spottom} shows a blow-up of the region around the
secondary star. The bright spot is shown close to the center of mass
of the late-type star, slightly located towards the leading
hemisphere. Since this is a velocity map and not a geometrical one,
there are at two possible interpretations of the position in space
of the bright spot (assuming the observed material is in the orbital
plane). The first one is that the emission is been produced at the
surface of the secondary, i.e. still attached to its gravitational
field. The second is that the emission is the result of a direct
shock front with the accretion disc and that the compact spot is
starting to gain velocity towards the Keplerian trajectory. We
believe that the second explanation is more plausible, as it is
consistent with the well accepted mechanism to produce a bright
spot. On the other hand, at this peculiar low state it is difficult
to invoke an external source strong enough to produce a
back-illuminated secondary and especially a bright and compact spot
on its leading hemisphere.

\section[]{Basic system parameters} \label{baspar}

Assuming that the radial velocity semi-amplitudes reflect accurately
the motion of the binary components, then from our results -- $K_{em}
= K_1 = 107 \pm 2$  km~s$^{-1}$; $K_{abs} = K_2 = 310 \pm 5$
km~s$^{-1}$ -- and adopting $P=0.1769061911$ we obtain:

$$q = {K_1 \over K_2} = {M_2 \over M_1} = {0.35 \pm 0.05},$$

$$M_1 \sin^3 i = {P K_2 (K_1 + K_2)^2 \over 2 \pi G} = 0.99 \pm 0.03 M_{\odot},$$

$$M_2 \sin^3 i = {P K_1 (K_1 + K_2)^2 \over 2 \pi G} = 0.35 \pm 0.02 M_{\odot},$$

and

$$ a \sin i = {P (K_1 + K_2) \over 2 \pi} = 1.46 \pm 0.02 R_{\odot}.$$

Using the inclination angle derived by \citet{zha87}, $i =69.7^\circ
\pm 0.7$, the system parameters become: $ M_{WD} = 1.20 \pm 0.05 \,
M_{\odot}$; $ M_{RD} = 0.42 \pm 0.04 \, M_{\odot}$; and $ a = 1.55
\pm 0.02 \, R_{\odot}$.

\subsection[]{The inner and outer size of the disc} \label{discsize}

A first order estimate of the dimensions of the disc -- the inner
and outer radius -- can be made from the observed Balmer emission
line. Its peak-to peak velocity separation is related to the outer
radius of the accreted material, while the wings of the line, coming
from the high velocity regions of the disc, can give an estimate of
the inner radius \citep{sma01}. The peak-to-peak velocity separation
of the 31 individual spectra were measured (see
section~\ref{double-peaks}), as well as the velocity of the blue and
red wings of $H\alpha$ at ten percent level of the continuum level.
>From these measurements we derive mean values of $V_{out} = 460 \,
\, {\rm km \, s^{-1}}$ and $V_{in} = 1200 \, \, {\rm km \, s^{-1}}$.

These velocities can be related to the disc radii from numerical
disc simulations, tidal limitations and analytical approximations
(see \citet{war95} and references therein). If we assume the
material in the disc at radius $r$ is moving with Keplerian
rotational velocity $V(r)$, then the radius in units of the binary
separation is given by \citep{hor86}:

$$ r/a = (K_{em} + K_{abs}) K_{abs} /V(r)^2,$$

The observed maximum intensity of the double-peak emission in
Keplerian discs occurs close to the velocity of its outer radius
\citep{sma81}. From the observed $V_{out}$ and $V_{in}$ values we
obtain an outer radius of $ R_{out}/a = 0.61$ and an inner radius of
$R_{in}/a = 0.09$. If we take $ a = 1.55 \pm 0.02 \, R_{\odot}$ from
the last section we obtain an inner radius of the disc
$R_{in}=0.1395 R_{\odot}$ equivalent to about 97\,000~km. This is
about 25 times larger than the expected radius of the white dwarf
(see section~\ref{discus}). On the other hand, the distance from the
center of the primary to the inner Lagrangian point, $R_{L_1}/a$, is

$$R_{L_1}/a = 1 - w + 1/3 w^2 + 1/9 W^3,$$

where $w^3= q/(3(1 + q)$ (\citep{kop59}). Using $q=0.35$ we obtain
$R_{L_1}/a = 0.63$. The disc, therefore, appears to be large, almost
filling the Roche-Lobe of the primary, with the matter leaving the
secondary component through the $L_1$ point colliding with the disc
directly and producing the hot spot near this location.

\section{Discussion} \label{discus}

For the first time, a radial velocity semi-amplitude of the primary
component of U~Gem has been obtained in the visual spectral region,
which agrees with the value obtained from ultraviolet observations
by \citet{sio98} and \citet{lon99}. In a recent paper, \citet{und06}
present high-resolution spectroscopy around H$\alpha$ and H$\beta$
and conclude that they cannot recover the ultraviolet value for
$K_1$ to better than about 20 percent by any method. Although the
spectral resolution at H$\alpha$ of the instrument they used is only
a factor of two smaller than that of the one we used, the diagnostic
diagrams they obtain show a completely different behavior as
compared to those we present here, with best values for $K_{1}$ of
about 95 km s$^{-1}$ from H$\alpha$ and 150 km s$^{-1}$ from
H$\beta$ (see their Figures 13 and 14, respectively). We believe
that the disagreement with our result lies not in the quality of the
data or the measuring method, but in the distortion of the emission
lines due to the presence of a complex accretion disc at the time of
their observations, as the authors themselves suggest. Their Doppler
tomograms show emission at low velocity, close to the center of
mass, two distinct spots, a narrow component close to the $L_1$
point, and a broader and larger one between the Keplerian and the
ballistic trajectories. There is even evidence of a weak spiral
structure. In contrast, we have observed U~Gem during a favorable
stage, one in which the disc was fully symmetric, and the hot-spot
was narrow and near the inner Lagrangian point. This allowed us to
measure the real motion of the white dwarf by means of the
time-resolved behavior of the H$\alpha$ emission line.

Our highly consistent results for the systemic velocity derived from
the H$\alpha$ spot ($\gamma = 33 \,\pm$ 10 km~s$^{-1}$ and those
found from the different methods used for the radial velocity
analysis of the emission arising from the accretion disk (see
section~\ref{prim} and Table~\ref{OrbParam}), give strong support to
our adopting a true systemic velocity value of $\gamma = 34 \,\pm$ 2
km~s$^{-1}$. If we are indeed detecting the true motion of the white
dwarf, we can use this adopted value, to make an independent check
on the mass of the primary: The observed total redshift of the white
dwarf (gravitational plus systemic)-- found by \citet{lon99} -- is
172~km~s$^{-1}$, from which, after subtraction of the adopted
systemic velocity, we derive a gravitational shift of the white
dwarf of 138~km s$^{-1}$. From the mass-radius relationship for
white dwarfs \citep{and88}, we obtain consistent results for $M_{wd}
= 1.23 M_{\odot}$ and $R_{wd}$ = 3900 km (see Figure 7 in
\citep{lon99}). This mass is in excellent agreement with that
obtained in this paper from the radial velocity analysis.

>From our new method to determine the radial velocity curve of the
secondary (section~\ref{ugemb}), we obtain a value for the
semi-amplitude close to 310 km s$^{-1}$. Three previous  papers have
determinations of the radial velocity curves from the observed Na~I
doublet in the near-infrared. In order to evaluate if our method is
valid, we here compare our result with these direct determinations.
The published values are: $ K_{rd} = 283$ km s$^{-1} \pm 15$
(\citep{wad81}); $K_{rd} = 309$ km s$^{-1} \, \pm 3$, (before
correction for irradiation effects, \citep{fri90}); and $ K_{rd} =
300$ km s$^{-1}$ \citep{nay05}. \citet{wad81} notes that an
elliptical orbital ($e=0.086$) may better fit his data, as the
velocity extremum near phase 0.25 appears somewhat sharper than that
near phase 0.75 (see his Figure 3). However, he also finds a very
large systemic velocity, $\gamma = 85$ km s$^{-1}$, much larger than
the values found by \citet{kra62} ($\gamma = 42$ km s$^{-1}$) and
\citet{sma76} ($\gamma = 40 \,\pm$ 6 km s$^{-1}$), both obtained
from the emission lines. Since the discrepancy with the results of
these two authors was large, \citet{wad81} defers this discussion to
further confirmation of his results. Instead, and more important,
this author discusses two scenarios that may significantly alter the
real value of $K_{2}$: the non-sphericity and the back-illumination
of the secondary. In the latter effect, each particular absorption
line may move further away from, or closer to the center of mass of
the binary. He estimates the magnitude of this effect and concludes
that the deviation of the photocenter would probably be much less
than 0.1 radii. \citet{fri90} further discusses the circumstances
that might cause the photocenter to deviate, and concludes that
their observed value for the semi-amplitude should be corrected down
by 3.5 percent, to yield $ K_{2} = 298$ km s$^{-1} \pm 9$. Although
they discuss the results by \citet{mart88} -- which indicate that
the relatively small heating effects in quiescent dwarf novae always
lead to a decrease in the measured $K_{rd}$ for the Na~I lines --
they argue that line quenching, produced by ionization of the same
lines, may also be important, and result in an increased $K_{rd}$.
Another disturbing effect, considered by the same authors, is line
contamination by the presence of weak disc features, like the
Paschen lines. In this respect we point out here that a poor
correction for telluric lines will function as an anchor, reducing
also the amplitude of the radial velocity measurements.
\citet{fri90} also find an observed systemic velocity of $\gamma =
43 \,\pm$ 6 km s$^{-1}$ and a small eccentricity of $e=0.027$.
\citet{nay05} also discuss the distortion effects on the Na~I lines
and, based on their fit residuals, argue in favor of a depletion of
the doublet in the leading hemisphere of the secondary, around
phases 0.4 and 0.6, as removing flux from the blueward wing of the
lines results in an apparent redshift, which would explain the
observed residuals. However, they additionally find that fitting the
data to an eccentric orbit, with $e=0.024$, results in a significant
decrease in the residuals caused by this depletion, and conclude
that it may be unnecessary to further correct the radial velocity
curve. We must point out that a depletion of the blueward wing of
the Na~I lines will results in a contraction of the observed radial
velocity curves, as the measured velocities -- especially around
phases 0.25 and 0.75 -- will be pulled towards the systemic
velocity. \citet{nay05} present their results derived from the Na~I
doublet and the K\,I/TiO region (around 7550-7750 \AA), compared
with several spectral standards, all giving values between 289 and
305 km s$^{-1}$ (no errors are quoted). Based on the radial velocity
measurements for Na~I, obtained by these authors in 2001 January
(115 spectra), and using GJ213 as template (see their Table 1), we
have recalculated the circular orbital parameters through our
nonlinear least-squares fit. We find $K_{2} = 300$ km s$^{-1} \, \pm
1$, in close agreement with their published value.

It would be advisable to establish a link between the observed gamma
velocity of the secondary and the semi-amplitude $K_2$, under the
assumption that its value may be distorted by heating effects. We
take as a reference our results from the radial velocity analysis of
the broad $H\alpha$ line and the hot-spot from the secondary, which
support a true systemic velocity of 34 km s$^{-1}$. However, we find
no positive correlation in the available results derived from the
Na~I lines, either between different authors or even among one data
set. In the case of \citet{nay05}, the gamma values show a range
between 11 and 43 km s$^{-1}$, depending on the standard star used
as a template, for $K_2$ velocities in the range 289 to 305 km
s$^{-1}$. \citet{wad81} finds $\gamma = 85 \,\pm$ 10 km s$^{-1}$ for
a low $K_2)$ value of 283 km s$^{-1}$, while \citet{fri90} finds
$\gamma = 43 \,\pm$ 6 km s$^{-1}$ for $K_2$ about 309 km s$^{-1}$,
and we obtain a large gamma velocity of about 69 km s$^{-1}$ for a
$K_2$ value of 310 km s$^{-1}$. We believe that further and more
specific spectroscopic observations of the secondary star should be
conducted in order to understand the possible distortion effects on
lines like the Na~I doublet, and their implications on the derived
semi-amplitude and systemic velocity values.

\section*{Acknowledgments}

E. de la F wishes to thank Andr\'es Rodriguez J. for his useful
computer help. The Thomson detector, used in our observations, was
obtained through PACIME-CONACYT project F325-E9211.

\label{lastpage}

\end{document}